\documentclass[twocolumn,English,aps,pra,10pt,superscriptaddress,floatfix]{revtex4-2}
\usepackage{times}
\usepackage{graphicx}
\usepackage{amssymb}
\usepackage{bm}% bold math
\usepackage{amssymb,amsfonts,amsmath,amsbsy,bm,t1enc,latexsym}
\usepackage{float}
\usepackage[colorlinks=true,citecolor=blue,linkcolor=magenta]{hyperref}
\usepackage[markup=blue, authormarkupposition=left]{changes}
\usepackage{url}
\usepackage[mode=text]{siunitx}
\DeclareSIUnit \dbc {dBc}
\DeclareSIUnit \dbm {dBm}
\DeclareSIQualifier\peak{p}
\usepackage{soul}
\usepackage{changes}
\usepackage{cancel}
\usepackage{times}
\usepackage{chemformula,xspace} % proper typesetting of chemical formula; smart blank space insertion for macros
\usepackage{braket}
\usepackage[capitalise]{cleveref} % for smart referencing; \Cref at the beginning of the sentence and \cref otherwise

\newcommand{\LN}[0]{\ch{LiNbO3}\xspace} 
\newcommand{\LT}[0]{\ch{LiTaO3}\xspace}
\begin{document}

%\title{Ultrabroadband Lithium Tantalate Electro-Optic Comb Generator via Monolithic Microwave Co-design}

%\title{\hl{JZ,discussion:}Co-designing microwave and photoncis integrated circuits for ultrabraod band electro-optic comb generation}

% \title{Ultra-broadband low power Lithium Tantalate Electro-Optic Frequency comb}
%\title{\hl{new suggestion, TJK? A bit long.} Ultra-broadband integrated electro-optical frequency combs using microwave integrated circuits}
%\title{\hl{new suggestion, TJK} Ultra-broadband electro-optical frequency combs using photonic and microwave co-integrated circuits}
%\title{\hl{new suggestion, TJK}Ultrabroadband electro-optical frequency comb using co-designed monolithic microwave and electro-optical integrated circuits}
\title{Integrated Triply Resonant Electro-Optic Frequency Comb in Lithium Tantalate} %Go ArXiv
%\title{Ultra-broadband Integrated Electro-Optic Frequency Comb}
%\title{Ultra-broadband and turn-key Electro-Optic Frequency Comb Generation in Lithium Tantalate with Monolithic Microwave Co-design}
%\title{Ultrabroadband triply resonant Lithium Tantalate electro-optical frequency comb }

\author{Junyin Zhang}\thanks{These authors contributed equally.}
\affiliation{Institute of Physics, Swiss Federal Institute of Technology Lausanne (EPFL), CH-1015 Lausanne, Switzerland}
\affiliation{Center of Quantum Science and Engineering, EPFL, CH-1015 Lausanne, Switzerland}

\author{Chengli Wang}\thanks{These authors contributed equally.}
\affiliation{Institute of Physics, Swiss Federal Institute of Technology Lausanne (EPFL), CH-1015 Lausanne, Switzerland}
\affiliation{Center of Quantum Science and Engineering, EPFL, CH-1015 Lausanne, Switzerland}

\author{Connor Denney}\thanks{These authors contributed equally.}
\affiliation{Departement of Electrical Engineering, Colorado School of Mines, Golden, Colorado 80401, United States}

\author{Grigory Lihachev}
\affiliation{Institute of Physics, Swiss Federal Institute of Technology Lausanne (EPFL), CH-1015 Lausanne, Switzerland}
\affiliation{Center of Quantum Science and Engineering, EPFL, CH-1015 Lausanne, Switzerland}

\author{Jianqi Hu}
\affiliation{Institute of Physics, Swiss Federal Institute of Technology Lausanne (EPFL), CH-1015 Lausanne, Switzerland}
\affiliation{Center of Quantum Science and Engineering, EPFL, CH-1015 Lausanne, Switzerland}

\author{Wil Kao}
\affiliation{Institute of Physics, Swiss Federal Institute of Technology Lausanne (EPFL), CH-1015 Lausanne, Switzerland}
\affiliation{Center of Quantum Science and Engineering, EPFL, CH-1015 Lausanne, Switzerland}

\author{Terence Blésin}
\affiliation{Institute of Physics, Swiss Federal Institute of Technology Lausanne (EPFL), CH-1015 Lausanne, Switzerland}
\affiliation{Center of Quantum Science and Engineering, EPFL, CH-1015 Lausanne, Switzerland}

\author{Nikolai Kuznetsov}
\affiliation{Institute of Physics, Swiss Federal Institute of Technology Lausanne (EPFL), CH-1015 Lausanne, Switzerland}
\affiliation{Center of Quantum Science and Engineering, EPFL, CH-1015 Lausanne, Switzerland}

\author{Zihan Li}
\affiliation{Institute of Physics, Swiss Federal Institute of Technology Lausanne (EPFL), CH-1015 Lausanne, Switzerland}
\affiliation{Center of Quantum Science and Engineering, EPFL, CH-1015 Lausanne, Switzerland}

\author{Mikhail Churaev}
\affiliation{Institute of Physics, Swiss Federal Institute of Technology Lausanne (EPFL), CH-1015 Lausanne, Switzerland}
\affiliation{Center of Quantum Science and Engineering, EPFL, CH-1015 Lausanne, Switzerland}

\author{Xin Ou}
\affiliation{State Key Laboratory of Materials for Integrated Circuits, Shanghai Institute of Microsystem and Information Technology, Chinese Academy of Sciences, Shanghai, China}

\author{Johann Riemensberger}
\email[]{johann.riemensberger@ntnu.no}
\affiliation{Norwegian University of Science and Technology, NTNU, NO-7491 Trondheim, Norway}

\author{Gabriel Santamaria-Botello}
\email[]{gabriel.santamariabotello@mines.edu}
\affiliation{Departement of Electrical Engineering, Colorado School of Mines, Golden, Colorado 80401, United States}

\author{Tobias J. Kippenberg}
\email[]{tobias.kippenberg@epfl.ch}
\affiliation{Institute of Physics, Swiss Federal Institute of Technology Lausanne (EPFL), CH-1015 Lausanne, Switzerland}
\affiliation{Center of Quantum Science and Engineering, EPFL, CH-1015 Lausanne, Switzerland}

\maketitle

%\section*{Abstract}
\textbf{
Integrated frequency comb generators based on Kerr parametric oscillation \cite{kippenberg2018dissipative} have led to chip-scale, gigahertz-spaced combs with new applications spanning hyperscale telecommunications, low-noise microwave synthesis, LiDAR and astrophysical spectrometer calibration \cite{marin2017microresonator,spencer2018optical,obrzudMicrophotonicAstrocomb2019,suh2019searching,riemensberger2020massively}. 
% \hl{Ref vahala and Herr on Astrocombs missing}
Recent progress in lithium niobate (\LN) photonic integrated circuits (PICs) has resulted in chip-scale electro-optic (EO) frequency combs \cite{zhang2019broadband,hu2022high}, offering precise comb-line positioning and simple operation without relying on the formation of dissipative Kerr solitons.
However, current integrated EO combs face limited spectral coverage due to the large microwave power required to drive the non-resonant capacitive electrodes and the strong intrinsic birefringence of \LN.
Here, we overcome both challenges with an integrated triply resonant architecture, combining monolithic microwave integrated circuits (MMICs) with PICs based on the recently emerged thin-film lithium tantalate (\LT) \cite{wang2023lithium}. 
With resonantly enhanced EO interaction and reduced birefringence in \LT, we achieve a four-fold comb span extension and a 16-fold power reduction compared to the conventional non-resonant microwave design.
Driven by a hybrid-integrated laser diode,
%and with less than \SI{7}{\watt} of on-chip power, 
the comb spans over \SI{450}{\nano\meter} ($>\SI{60}{\tera\hertz}$) with $> 2000$ lines, and the generator fits within a compact \SI{1}{\square\centi\meter} footprint.
We additionally observe that the strong EO coupling leads to an increased comb existence range approaching the full free spectral range of the optical microresonator.
The ultra-broadband comb generator, combined with detuning-agnostic operation,
could advance chip-scale spectrometry and ultra-low-noise millimeter wave synthesis \cite{li2014electro,sun2024integrated,kudelin2024photonic,zhao2024all} and unlock octave-spanning EO combs.
The methodology of co-designing microwave and optical resonators can be extended to a wide range of integrated electro-optics applications \cite{yu2022integrated, hu2021chip, yu2023integrated}.
}
%%%%%%%%%%%%%%%%%%%%%%%%%%%%%%%%%%%%%%%%%%%%%%%%%%%%%%%%%%%%%%%%%%%%%%%%%%%%%%%%%%%%%%%%%%%%%%%%%%%%%%%%%%%%%%%%%%%%%
%%%%%%%%%%%%%%%%%%%%%%%%%%%%%%%%%%%%%%%%%%%%%%%%%% Main Content starts here %%%%%%%%%%%%%%%%%%%%%%%%%%%%%%%%%%%%%%%%%
%%%%%%%%%%%%%%%%%%%%%%%%%%%%%%%%%%%%%%%%%%%%%%%%%%%%%%%%%%%%%%%%%%%%%%%%%%%%%%%%%%%%%%%%%%%%%%%%%%%%%%%%%%%%%%%%%%%%%
%\hl{Introduction.}
%%%%%%%%%%%%%%%%%%%%%%%%%%%%%%%%%%%%%%%%%%%%%%%%%%%%%%%%%%%%%%%%%%%%%%%%%%%%%%%%%%%%%%%%%%%%%%%%%%%%%%%%%%%%%%%%%%%%%%
%% Fig 1
\begin{figure*}[htbp]
	\centering
	\includegraphics[width=0.8\linewidth]{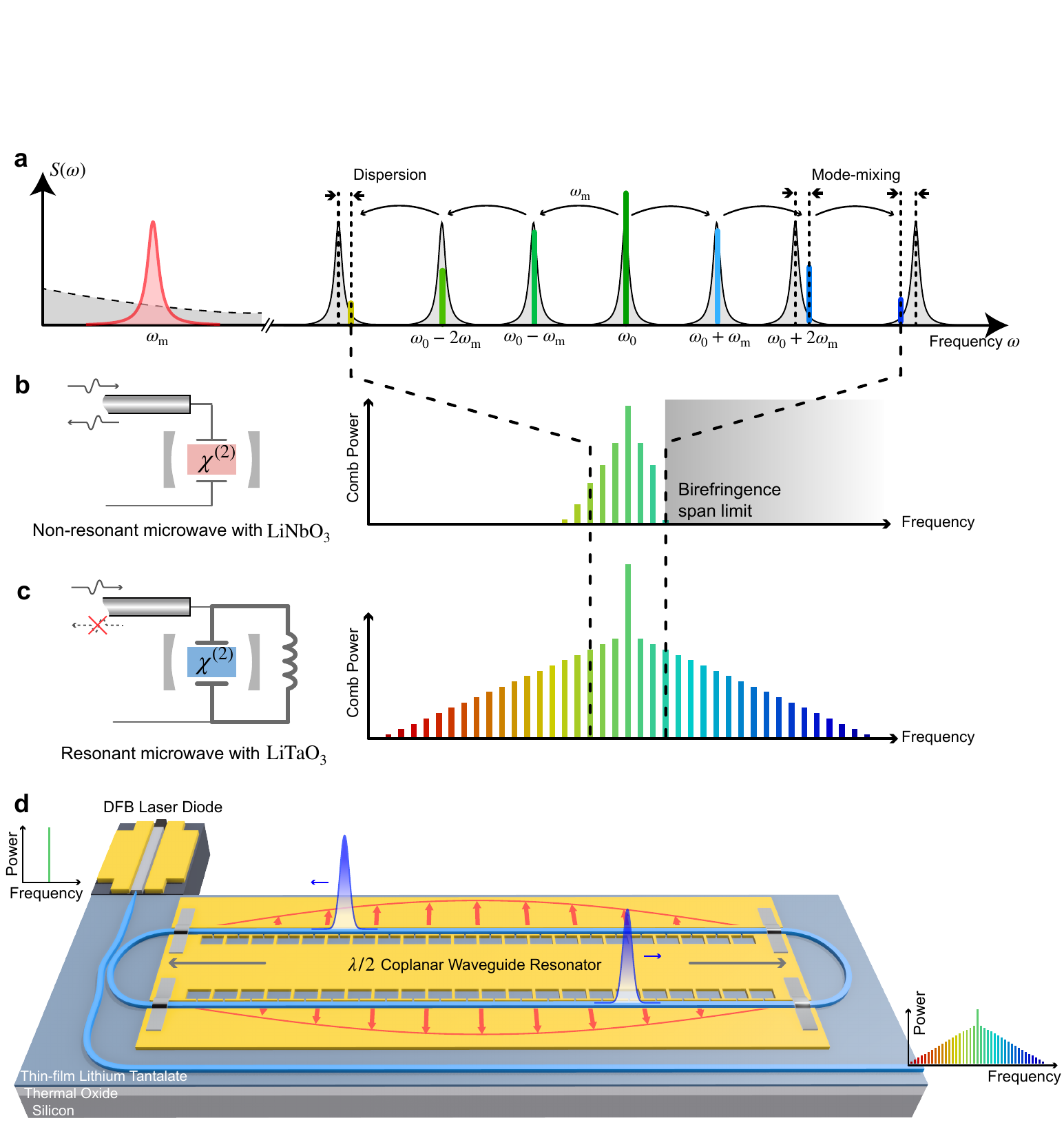}
	\caption{\textbf{Frequency comb generator based on co-designed monolithic microwave and electro-optic integrated circuits.} Triply-resonant lithium tantalate electro-optic comb generator.
	(a) Principle of electro-optic (EO) frequency comb generation and limitations in conventional implementations. The optical modes are separated by a nearly constant free spectral range (FSR). A strong microwave pump with frequency $\omega_\mathrm{m}$ matching the FSR drives cascaded frequency conversion processes via the Pockels effect. Optical dispersion and mode-mixing can distort the evenly spaced optical resonances, limiting sideband generation efficiency away from the optical input mode. 
	(b) Schematic of non-resonant electrode design for existing integrated comb generators. The commonly employed \LN results in birefringent mode mixing that limits the comb span.
    (c) Schematic of the present triply resonant architecture. The significantly weaker birefringence in \LT suppresses mode mixing.
    (d) Schematic of the hybrid integrated comb generator with a dispersion-engineered \LT racetrack microresonator and monolithic microwave integrated $\lambda/2$ co-planar waveguide resonator. 
    The enhanced EO interaction reduces the sensitivity of broadband comb formation to pump laser detuning, enabling robust turnkey operation with a hybrid integrated distributed feedback (DFB) diode laser.
	}
	\label{fig1}
\end{figure*}

Microresonator optical frequency combs utilizing ultra-low loss and wafer-scale manufacturable photonic integrated circuits (PICs)---especially those based on foundry-available silicon nitride \cite{kippenberg2011microresonator}---have been pivotal in advancing their fiber-based laboratory counterparts to chip-scale system-level applications in science and technology.
Their versatility has been demonstrated in hyper-scale data communication \cite{marin2017microresonator}, parallel LiDAR \cite{riemensberger2020massively}, neuromorphic computing \cite{feldmann2021parallel}, ultra-low-noise microwave synthesis \cite{zhao2024all,kudelin2024photonic,sun2024integrated}, broadband spectroscopy \cite{suh2016microresonator}, and astrophysical spectrometer calibration \cite{obrzudMicrophotonicAstrocomb2019,suh2019searching}.

The availability of thin-film lithium niobate (\LN) using smart-cut \cite{levy1998fabrication} has triggered the development of electro-optic photonic integrated circuits with a large Pockels coefficient \cite{zhu2021integrated, wang2018integrated}. 
{This platform has re-ignited interest in electro-optic (EO) frequency combs \cite{kourogiWidespanOpticalFrequency1993,rueda2019resonant}.} 
The recently emerged integrated EO combs \cite{zhang2019broadband,hu2022high} complement soliton microcombs, exhibit similar compactness while offering innate stability of the repetition rates set by the microwave modulation frequency.
Coherent sideband generation mediated by the EO Pockels effect does not have a minimum optical threshold, unlike parametric oscillations in Kerr comb formation. 
Additionally, it does not require the complex laser tuning mechanisms needed for dissipative Kerr soliton initiation \cite{herrTemporalSolitonsOptical2014,yiActiveCaptureStabilization2016,lucasDetuningdependentPropertiesDispersioninduced2017,guoUniversalDynamicsDeterministic2017,stoneThermalNonlinearDissipativeSoliton2018}.
Microcombs based on dissipative Kerr solitons also suffer from reduced conversion efficiency at lower repetition rates, particularly in the sub-\SI[number-unit-product={\text{-}}]{100}{\giga\hertz} frequencies (e.g., X-band for radar, K-band for 5G) \cite{liu2020photonic}.

Despite these advantages, EO combs still face outstanding challenges. The achieved frequency comb span, along with the line count, has been limited compared to soliton microcombs, which have attained octave spanning operation \cite{pfeiffer2017octave,spencer2018optical} and the generation of more than 2000 comb lines \cite{anderson2021photonic,cheng2024frequency}.
This is due to the insufficient EO coupling rate for generating thousands of sidebands \cite{rueda2019resonant,zhang2019broadband,hu2022high}.
As a result, large microwave pump power is needed to attain the requisite modulation depth. 
State-of-the-art integrated EO combs therefore require specialized power microwave circuits and bulk protective circulators \cite{zhang2019broadband,hu2022high}, which remain challenging to integrate into chip-scale systems.
A further limitation stems from the intrinsic birefringence of \LN that imposes a span limit due to mode-mixing \cite{hu2022high, hu2022mirror}. 
Together, these factors have limited the bandwidth of state-of-the-art integrated EO comb below \SI{140}{nm} and exacerbated their microwave power requirements  \cite{zhang2019broadband,hu2022high}. %JZ: power is bit tricky, I don't think the power reported in Loncar's paper is correct.
Here, we overcome these challenges by bringing coplanar waveguide resonators from monolithic microwave integrated circuits (MMICs) into photonic integrated circuits and implementing an integrated triply resonant EO comb generator. The tight field confinement offered by microwave-photonic co-integration enhances the single-photon EO coupling rate by more than 300 times compared to bulk implementations \cite{rueda2019resonant}. 
Combined with dispersion-engineered lithium tantalate (\LT) photonic integrated circuits that exhibit 17 times lower intrinsic birefringence than the workhorse EO material \LN, and driven by a hybrid integrated semiconductor laser diode, the device is capable of generating over 2000 sidebands (a \SI[number-unit-product={\text{-}}]{450}{\nano\meter} span) while consuming under \SI{7}{\watt} of on-chip power. We attain an \SI[number-unit-product={\text{-}}]{80}{\nano\meter} span with only \SI{13}{\dbm} of microwave power ($<\SI{1.5}{\volt\peak}$), representing an over 16-fold power reduction compared to a conventional non-resonant electrode design.
Moreover, the enhanced EO coupling rate in our triply resonant scheme is shown to lead to unprecedented detuning-agnostic operation with a comb existence range exceeding \SI{90}{\percent} of the free spectral range (FSR) despite utilizing a microresonator, enabling full FSR sweeping of the comb lines free of spectral holes.

\section*{Results}
\subsection*{Triply resonant cavity electro-optic architecture}
We leverage resonantly enhanced cascaded energy transfer, illustrated in \cref{fig1}a, for efficient EO comb generation, one of the earliest studied methods to generated optical frequency combs \cite{kourogiWidespanOpticalFrequency1993}.
\Cref{fig1}b delineates the non-resonant lumped capacitive electrode design employed in state-of-the-art integrated EO combs based on \LN \cite{zhang2019broadband,hu2022high}.
As a result of the impedance mismatch with the input transmission line, a significant portion of the applied microwave power is reflected and dissipated as heat in the internal impedance of the microwave source.
This inefficiency in power delivery necessitates the use of RF amplifiers to broaden the comb, which has, to date, limited state-of-the-art integrated EO combs to a span below \SI{140}{\nano\meter} \cite{zhang2019broadband,hu2022high}.
Impedance mismatch can be overcome by employing a microwave cavity (\cref{fig1}c), which increases EO comb generation efficiency \cite{rueda2019resonant}. However, this method has only been implemented with bulk optical and microwave cavities rather than in integrated EO combs.
The present design, depicted in \cref{fig1}d, comprehensively tackles both challenges by integrating a monolithic microwave resonator with a \LT photonic racetrack resonator. 
Such triply resonant devices realized using bulk microwave cavities were originally considered for microwave photonic receivers \cite{ilchenkoSubmicrowattPhotonicMicrowave2002,ilchenkoWhisperinggallerymodeElectroopticModulator2003}, and more recently, have been investigated for quantum coherent transduction between the microwave and the optical domains \cite{ruedaEfficientMicrowaveOptical2016,javerzac-galyOnchipMicrowavetoopticalQuantum2016,fanSuperconductingCavityElectrooptics2018,holzgrafe2020cavity,mckenna2020cryogenic}. 
The deep sub-wavelength confinement of the microwave field of the co-planar waveguide resonator additionally leads to significantly enhanced EO coupling rates \cite{javerzac-galyOnchipMicrowavetoopticalQuantum2016}.
The advantage of this approach is seen by considering a three-wave mixing process between a pair of optical modes ($\hat{a}_{\mu}$, $\hat{a}_{\mu+1}$) and a microwave mode ($\hat{b}$) as in cavity electro-optics \cite{tsangCavityQuantumElectrooptics2010,ruedaEfficientMicrowaveOptical2016}, described by the Hamiltonian
\begin{equation}\label{eq_The_H}
\mathcal{\hat{H}} = \sum_{\mu} \hbar\Delta_\mu \hat{a}^\dagger_{\mu}\hat{a}_{\mu} - \sum_{\mu} \hbar g_0\left(\hat{a}^\dagger_{\mu+1} \hat{a}_{\mu} \hat{b} + \mathrm{h.c.}\right),
\end{equation}
where $\hbar$ denotes the reduced Planck's constant, and $\Delta_{\mu} = D_{\mathrm{int}}(\mu) + \Delta_\mathrm{L}$ is the gross detuning with contributions from optical integrated dispersion $D_{\mathrm{int}}$ and input laser detuning $\Delta_\mathrm{L}$. 
Intimately related to the internal generator performance, the single-photon EO coupling rate due to $\chi^{(2)}$ nonlinearity is given by
\begin{equation}
    g_0 \propto \sqrt{\frac{1}{V_\mu V_{\mu+1}V_\mathrm{m}}}\int_\text{\LT} \chi^{(2)}_{\alpha\beta\gamma} \Psi_{\mu,\alpha} \Psi_{\mu+1,\beta}^* \Psi_{\mathrm{m},\gamma}\,\mathrm{d}V.
\end{equation}
The mode volumes $V_k$ and $n^\text{th}$ components of the spatial field distribution function $\Psi_{k,n}$(\textbf{r}) together define the vacuum electric field components $E_{k,n}(\textbf{r}) = \sqrt{\hbar\omega_k/(2\epsilon_k V_k)}\Psi_{k,n}$.
The components are associated with mode $k$ at frequency $\omega_k$ and permittivity $\epsilon_k$, with $k = \{\mu,\mu+1,\mathrm{m}\}$ and $\mu$ as the optical longitudinal mode index.
Monolithic integration of the microwave and photonic subsystems enables reduced mode volumes and improved field overlap, allowing substantial increases in $g_0$.
The subsequently enhanced EO interaction internal to the device alleviates the constraint on on-chip microwave power. 
With a strong, undepleted microwave pump, the effective coupling rate between the optical modes is $g = g_0\braket{\hat{b}} = g_0\sqrt{n_\mathrm{m}}$, where $n_\mathrm{m}$ is the mean intracavity (microwave) photon number. 
%It can be shown that the comb span is proportional to $g$, and the external input microwave power required to generate a given comb span scales with $g_0^{-2}$ (Supplementary Information).
This coupling rate $g$ is proportional to the comb span (Supplementary Information). Compared to non-resonant electrode designs, we reduce the required power by incorporating 
a $\lambda/2$ CPW resonator, which enhances $n_\mathrm{m}$ by the resonator finesse, reducing the external input microwave power required to generate a given comb span.

%%%%%%%%%%%%%%%%%%%%%%%%%%%%%%%%%%%%%%%%%%%%%%%%%%%%%%%%%%%%%%%%%%%%%%%%%%%%%%%%%%%%%%%%%%%%%%%%%%%%%%%%%%%%%%%%%%%%%%
\subsection*{Birefringence span limit for electro-optic combs} %%%%%%%%%%%%%%%%%%%%%%%%%%%%%%%%%%%%%%%%%%%%%%%%%%%%%%%%%%%
%%%%%%%%%%%%%%%%%%%%%%%%%%%%%%%%%%%%%%%%%%%%%%%%%%%%%%%%%%%%%%%%%%%%%%%%%%%%%%%%%%%%%%%%%%%%%%%%%%%%%%%%%%%%%%%%%%%%%%
%% Fig 2
\begin{figure*}[htbp]
	\centering
	\includegraphics[width=0.8\linewidth]{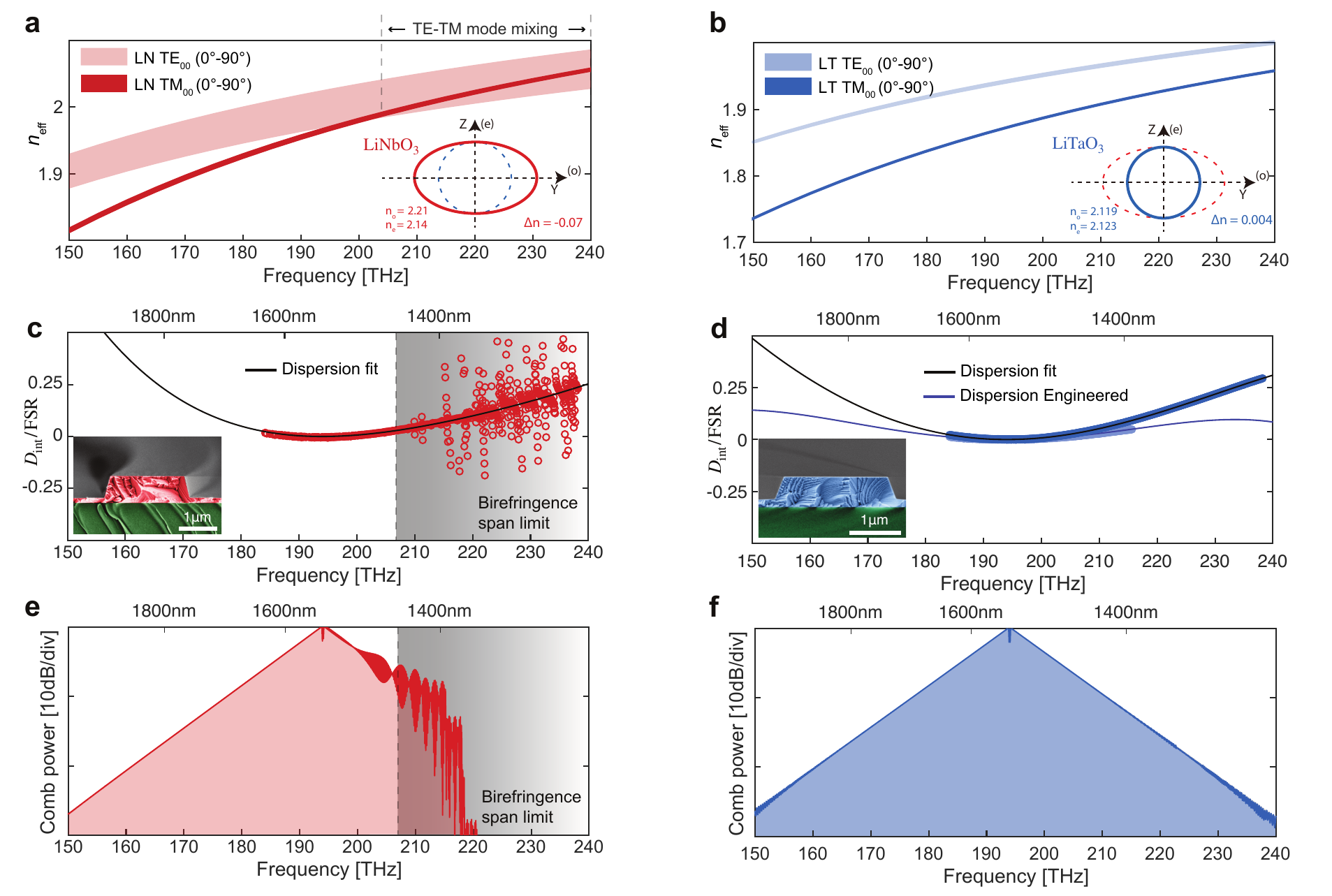}
	\caption{\textbf{Comparison of birefringent mode mixing effect and comb span for \LT and \LN microresonators}
		(a), (b) Simulated effective index for the fundamental quasi-transverse electric (TE) and magnetic (TM) modes in a \SI{2.0}{\micro\meter} $\times$ \SI{600}{\nano\meter} ridge waveguide with a \SI{100}{\nano\meter} slab in \LN (a) and \LT (b) in different waveguide directions. Inset shows \LN's (red) strong negative and \LT's (blue) weak positive uniaxial crystal birefringence.
		(c), (d) The measured dispersion profile for \LN (c) and \LT (d) with similar cross sections. In \LN, strong mode mixing occurs at frequencies above \SI{215}{\tera\hertz}, corresponding to the optical telecommunication E-band and O-band, in the \SI{2.0}{\micro\meter} wide waveguide. In contrast, mode mixing is not observed in both the \SI{2.0}{\micro\meter} wide and the dispersion-engineered \SI{2.1}{\micro\meter} wide waveguides in \LT. Insets depict cross-sectional scanning electron micrographs of the fabricated \LN and \LT photonics waveguides.
		(e), (f) Simulated comb spectra based on measured dispersion for \SI{2.0}{\micro\meter} wide \LN (e) and \LT (f) waveguides with an effective coupling rate $g = 0.35 \omega_{\mathrm{FSR}}$.
	}
\label{fig2}
\end{figure*}

In addition to enhancing EO interaction, realizing broadband comb generation involves managing the optical dispersion such that $\Delta_{\mu} \approx 0$ across the broadest possible wavelength range.
We first investigate the attainable dispersion in \LN and \LT racetrack resonators.
In microresonators based on x-cut \LN, transverse-electric (TE) modes experience a significant refractive index change from the ordinary ($n_{\mathrm{o}} = 2.21$) to extraordinary ($n_{\mathrm{e}}=2.14$) values across the waveguide bends, whereas the transverse-magnetic (TM) modes experience mostly $n_{\mathrm{o}}$.
As exemplified in \cref{fig2}a, the fundamental TE$_{00}$ mode thus crosses and hybridizes with the TM$_{00}$ mode, which 
% distorts the optical polarization in the waveguide bend and thus 
introduces strong intra-modal coupling \cite{pan2019fundamental}. 
The gigahertz-level mode crossings result in uneven frequency spacings between longitudinal modes of the microresonator, interrupting the coherent spectral extension of the comb towards shorter wavelengths. 
For our \SI{600}{\nano\meter} thick waveguide design, we find that the \LN resonance spectrum is distorted above a critical frequency of \SI{207}{\tera\hertz} (\SI{1450}{\nano\meter}). 
The critical frequency decreases with increasing waveguide thickness \cite{pan2019fundamental}, which limits the achievable flat dispersion span required for broadband comb generation in \LN racetrack resonators \cite{hu2022high,hu2022mirror}. 
As illustrated in \cref{fig2}b, by replacing \LN with \LT, which has 17 times lower birefringence \cite{wang2023lithium} (\( n_{\mathrm{e}} - n_{\mathrm{o}} = 0.004 \)), we suppress the crossing of TE and TM modes.
We compare the measured dispersion profiles of two identical racetrack resonators fabricated from \LN (\cref{fig2}c) and \LT (\cref{fig2}d). The results reveal significantly reduced local resonance frequency distortion due to birefringence mode mixing in \LT.
Furthermore, the dispersion-engineered \LT waveguide in \cref{fig2}d provides a flat global dispersion profile without measurable local birefringence distortion, making it suitable for octave-spanning EO comb generation.
\Cref{fig2}e and \ref{fig2}f compare the simulated comb spectra for \LN and \LT with an effective coupling rate $g = 0.35 \omega_{\mathrm{FSR}}$, using the non-dispersion-engineered $D_\mathrm{int}$ profiles from the measurement depicted in \cref{fig2}c and \ref{fig2}d. 
We observe that the comb line power in \LN drops significantly beyond \SI{1450}{\nano\meter} due to birefringence mode mixing, whereas the comb span in \LT is not limited by this issue.

%%%%%%%%%%%%%%%%%%%%%%%%%%%%%%%%%%%%%%%%%%%%%%%%%%%%%%%%%%%%%%%%%%%%%%%%%%%%%%%%%%%%%%%%%%%%%%%%%%%%%%%%%%%%%%%%%%%%%%
\subsection*{Device implementation} %%%%%%%%%%%%%%%%%%%%%%%%%%%%%%%%%%%%%%%%%%%%%%%%%%%%%%%%%%%
%%%%%%%%%%%%%%%%%%%%%%%%%%%%%%%%%%%%%%%%%%%%%%%%%%%%%%%%%%%%%%%%%%%%%%%%%%%%%%%%%%%%%%%%%%%%%%%%%%%%%%%%%%%%%%%%%%%%%%
\begin{figure*}[htbp]
	\centering
	\includegraphics[width=1\linewidth]{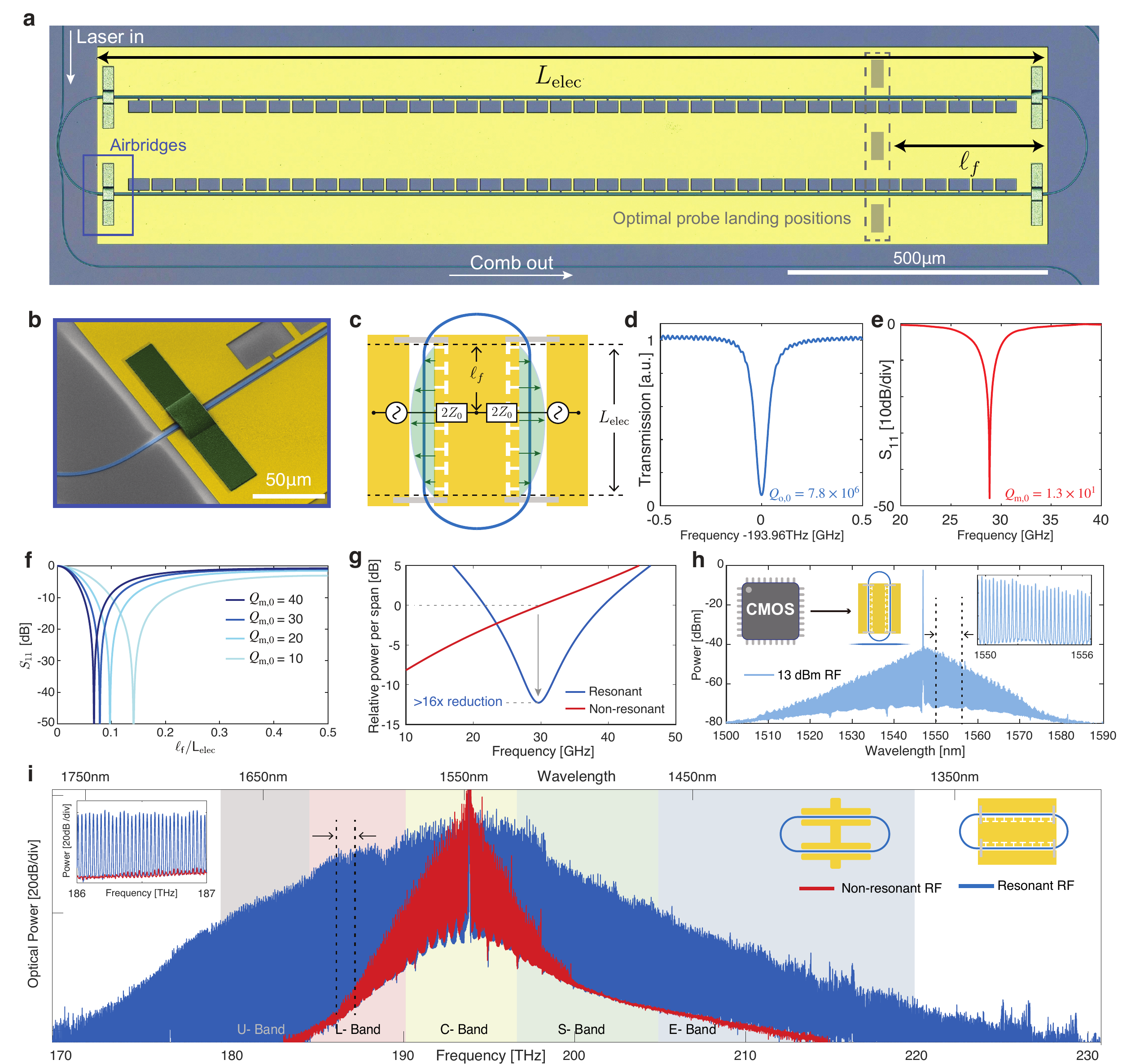}
	\caption{\textbf{Triply resonant electro-optic frequency comb with monolithic integrated microwave resonator.}
		(a) Optical micrograph of the integrated \LT electro-optic (EO) frequency comb generator. The comb generator consists of an air-cladded \LT racetrack microresonator (dark blue) and a coplanar waveguide (CPW) resonator (yellow) with aluminum air bridges forming a short-terminated $\lambda/2$ resonator. The microwave resonant frequency is engineered by inductive-loaded microstructures to match the optical repetition rate. 
		(b) Colorized SEM of the air bridge and the microstructured CPW. The etched \LT photonic waveguide (dark blue) passes underneath the air bridge.
        (c) Schematic of the microwave and photonic resonators. The external coupling rate of the CPW resonator can be tuned by changing the probe landing position $\ell_f$ continuously to achieve critical coupling.
        (d) Normalized transmission of \LT racetrack resonator, with an intrinsic quality factor $Q_\mathrm{o,0}$ exceeding seven million.
        (e) Measured reflection coefficient of the monolithic integrated CPW microwave resonator exhibiting $>\SI{40}{\decibel}$ resonant return loss. The resonance frequency matches the optical FSR (within \SI{9}{\decibel} bandwidth).	
		(f) The simulated microwave $S_{11}$ with different intrinsic microwave quality factor and probe landing positions $\ell_f$. The effective external coupling rate can be tuned by the probe landing position to achieve critical coupling for a wide range of intrinsic quality factors $Q_\mathrm{m,0}$.
    	(g) Simulated required microwave power for a given comb span with resonant and non-resonant electrode designs. 
		(h) Efficient comb generation with CMOS-level on-chip microwave pump power of \SI{13}{\dbm}. 
        (f) Measured output spectrum of the generated EO comb with conventional non-resonant electrodes (red) and with a monolithic CPW resonator (blue). The resonant design broadens the comb from \SI[number-unit-product={\text{-}}]{100}{\nano\meter} span to \SI[number-unit-product={\text{-}}]{450}{\nano\meter} span, with the slope reduced from %\SI{1.2}{\nano\meter\per\decibel} to \SI{6.0}{\nano\meter\per\decibel}. 
        \SI{1.2}{\decibel\per\tera\hertz} to \SI{0.3}{\decibel\per\tera\hertz}.
        The generated \LT EO comb spans the entire optical telecommunications L-, C-, S-, and E-bands, surpassing the birefringence limitations faced by \LN EO combs. Insets: magnified view of comb lines between \SI{186}{\tera\hertz} to \SI{187}{\tera\hertz}.
	}
\label{fig3}
\end{figure*}

To realize the integrated triply resonant architecture, we employ an optical \LT racetrack resonator embedded in a gold $\lambda/2$ coplanar-waveguide microwave resonator periodically loaded with an inductive slotted micro-structure, as illustrated in \cref{fig3}a.
This loading increases the intrinsic microwave quality factor $Q_{\mathrm{m},0}$ %reduce the ohmic loss 
\cite{shin2010conductor,kharel2021breaking} and simultaneously creates a slow-wave effect that aligns the resonance frequency with the optical FSR. 
Suspended metal bridges at both ends form short-circuit terminations (\cref{fig3}b), providing
maximum electric field strength at the center (\cref{fig3}c). 
This field distribution achieves theoretical optimal phase matching between microwave and optical modes, maximizing $g_0$ (Supplementary Information). The microwave pump is coupled through a ground-signal-ground microwave probe positioned off-center for critical coupling. 
The high-confinement \LT photonic waveguides allow for gold electrodes to be placed \SI{2}{\micro\meter} away from their edges while maintaining high intrinsic optical quality factors $> 7\times 10^6$ (\cref{fig3}d). 
\Cref{fig3}e shows the measured reflection coefficient $S_{11}$ of the critically coupled microwave resonator, with a $Q_{\mathrm{m},0}\approx 13$. This resonator topology allows for coupling rate tuning via adjustment of the probe landing position without affecting the field distribution. This technique guarantees a critically coupled microwave mode for a wide range of intrinsic quality factors, as illustrated in \cref{fig3}f. 
The optical comb sidebands and microwave pump are simultaneously resonant to increase the modulation efficiency. 

In our triply resonant system, the single-photon coupling rate is measured to be $g_0 = 2\pi\times\SI{2.2}{\kilo\hertz}$ (Supplementary Information). 
The microwave resonator is critically coupled to maximize the intracavity photon number $n_\mathrm{m}$ and consequently the effective coupling rate $g$. 
For a critically coupled $\lambda/2$ standing-wave resonator, the intracavity power is enhanced by its finesse %$P_{\mathrm{intra}}/P_\mathrm{m}$ is enhanced by a factor 
$\mathcal{F}=Q_{\mathrm{m},0}/(2\pi)\approx 2$ with respect to the incident power. 
A four-fold improvement in $g$ is expected compared to the modulation depth of a \SI{50}{\micro\meter} wide non-resonant electrode structure (Supplementary Information). 
The improvement is substantially higher than $\sqrt{\mathcal{F}}$ because the non-resonant electrodes do not behave as a system of unity finesse. Instead, waves are inefficiently coupled to the electrodes due to impedance mismatch with the probe, subsequently undergoing multiple reflections and partial destructive interference. The effect can be fully captured by a transmission line model (Supplementary Information).
This also implies that the resonant structure requires 16 times less microwave power to achieve the same span, as shown in \cref{fig3}g.
\Cref{fig3}h shows the triply resonant comb spectrum generated with \SI{13}{\dbm} of on-chip power on an under-coupled LTOI resonator. 
\Cref{fig3}i compares the EO comb spectrum generated in two similar over-coupled LTOI racetrack microresonators with non-resonant and resonant microwave structures, both featuring identical optical waveguide cross sections and waveguide-electrode gaps. In both cases, the devices are pumped with an on-chip microwave power of \SI{37}{\dbm} at a carrier frequency around \SI{29.6}{\giga\hertz}. 
In the non-resonant design, we measure a comb span of \SI{8}{\tera\hertz} at \SI{-20}{\decibel} from the first-order sidebands, corresponding to an approximate comp slope of \SI{1.2}{\decibel\per\tera\hertz}. 
A fourfold increase in the comb span (or 16 times lower power requirement for the same span) is observed as a result of the microwave resonator, with over \SI{450}{nm} measured span (\cref{fig3}e) at a slope of about \SI{0.3}{\decibel\per\tera\hertz}. 
The slight asymmetry of the comb is due to the frequency-dependent optical coupling rate.

\begin{figure*}[htbp]
	\centering
	\includegraphics[width=1\linewidth]{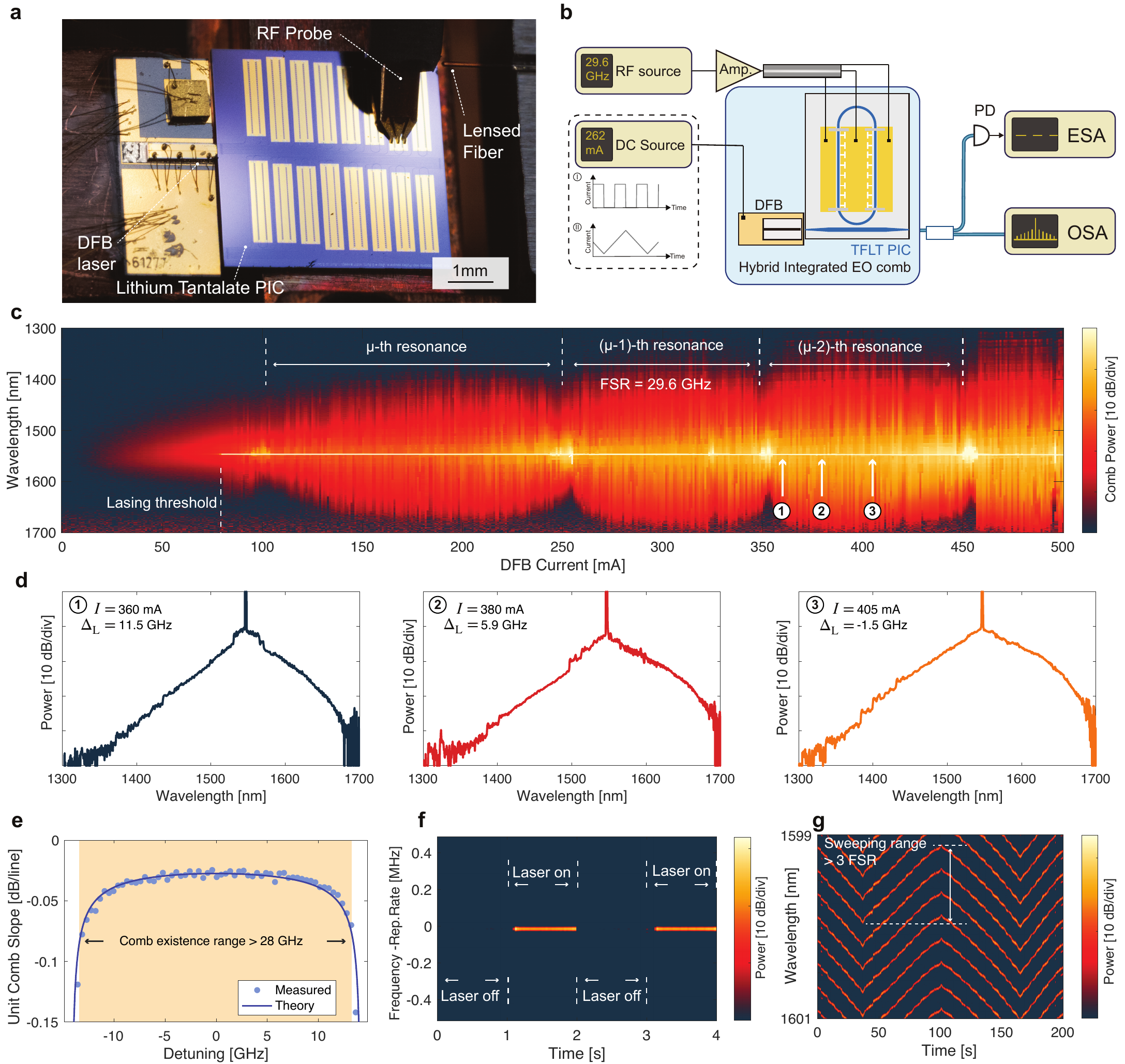}
	\caption{\textbf{Detuning-agnostic hybrid-integrated ultra-broadband comb source.}
	(a) Optical image of the hybrid integrated triply resonant electro-optic (EO) comb generator.
	(b) Experimental setup for EO comb generation and characterization. DFB: Distributed feedback diode laser. OSA: Optical spectrum analyzer. PD: Fast photodiode. AMP: Microwave power amplifier.
    (c) Optical spectrogram of the generated EO comb during laser diode current ramping. The comb is immediately generated when the DFB diode starts lasing, and the broadband comb can be maintained in a wide current range (corresponding to pump laser detuning) across three optical free spectral ranges (FSR). 
	(d) The optical spectra---line cuts at the indicated DFB currents in (c)---with several laser diode current settings. Each corresponding laser-cavity detuning, $\Delta_\mathrm{L}$, is labeled. 
	(e) Measured normalized comb slope $P_{\mu+1}/P_{\mu}$ at \SI{10}{\tera\hertz} offset (\SI{1640}{\nano\meter}) from the pump as a function of laser detuning. 
    The comb existence range exceeds \SI{28}{\giga\hertz}, covering more than \SI{90}{\percent} of the free spectral range (FSR). 
    Within this range, the comb shape remains nearly unchanged.
	(f) Comb repetition rate signal detected during periodic laser on-off switching, demonstrating stable turnkey operation.
	(g) Time-wavelength spectrogram of the measured comb line positions at \SI{7}{\tera\hertz} offset when diode current undergoes triangle-wave modulation. 
    Sweeping across an entire FSR is achievable due to the microwave enhancement. 
	}
\label{fig4}
\end{figure*}

Next, we study the range of laser detuning over which EO combs can be sustained, in the presence of the MMIC cavity-enhanced microwave field.
\Cref{fig4}a shows a photograph of the hybrid-integrated comb generator, and \cref{fig4}b depicts the experimental setup.
We butt-couple a distributed feedback (DFB) laser diode to the edge of the LTOI chip with a coupling loss of \SI{3}{\decibel} at the facet. The laser diode is controlled by an external DC current source, and the generated comb is collected from the photonic chip with a lensed fiber. 
\Cref{fig4}c illustrates the variations in comb spectra for different laser currents, and three example spectra are presented in \cref{fig4}d. As the current increases, the pump laser frequency is swept across three FSRs. Notably, we obtain a broadband comb immediately after the current passes the lasing threshold, and the broad span can be maintained over a wide current range. 
The comb existence range is given by $ |\Delta_\mathrm{L}| < 2g $ (Supplementary Information).  
\Cref{fig4}e depicts the measured comb slope at a \SI[number-unit-product={\text{-}}]{10}{\tera\hertz} offset (\SI{1630}{\nano\meter}) from the laser carrier ($\approx \SI{1547}{\nano\meter}$) as a function of laser detuning. The comb slope remained nearly constant for over \SI{28}{\giga\hertz}, which constitutes $90\%$ of the FSR. This observation indicates that for a randomly chosen pump frequency, a comb with a similar span can be generated with $90\%$ probability. 
\Cref{fig4}f illustrates the repetition rate signal measured during periodic laser switching, demonstrating stable turnkey operation.
The large comb existence range enables continuous sweeping of all comb lines across the entire FSR, leaving no spectral hole within the span (\cref{fig4}g). This feature is particularly crucial for chip-based sensing \cite{suh2016microresonator} and coherent ranging applications \cite{riemensberger2020massively}.

\section*{Conclusion and outlook}%%%%%%%%%%%%%%%%%%%%%%%%%%%%%%%%%%%%%%%%%%%%%%%%%%%%%%%%%%%
%%%%%%%%%%%%%%%%%%%%%%%%%%%%%%%%%%%%%%%%%%%%%%%%%%%%%%%%%%%%%%%%%%%%%%%%%%%%%%%%%%%%%%%%%%%%%%%%%%%%%%%%%%%%%%%%%%%%%%
In summary, we have demonstrated an ultra-broadband integrated triply resonant EO frequency comb generator using the newly emerged thin-film \LT platform \cite{wang2023lithium}. 
The material choice helps overcome the birefringence span limit for spectral coverage faced by the conventionally used \LN \cite{zhang2019broadband,hu2022high,rueda2019resonant}. 
We utilize a monolithic standing-wave microwave resonator to optimize for field overlap and phase matching with the optical modes, achieving a single-photon EO coupling rate $g_0 = 2\pi\times\SI{2.2}{\kilo\hertz}$, over 300 times larger than prior bulk resonant realization \cite{rueda2019resonant}. 
This critically coupled MMIC resonator additionally enhances the intracavity microwave pump photon number while eliminating the power reflection, and, by extension, the effective EO coupling rate.
Strong coupling enables broadband comb generation over a laser detuning range exceeding \SI{90}{\percent} FSR, facilitating turnkey operation with a free-running DFB laser diode and full FSR sweeping.
This feature permits hybrid integration of the laser diode such that the comb generator fits within a \SI{1}{\square\centi\meter} footprint. 
% The field enhancement and reflection reduction from the monolithic microwave resonator enable an ultra-broad electro-optic comb within this small footprint with gross power below \SI{7}{\watt}.
Crucially, as a result of the efficient optical sideband generation, we attain a comb span greater than \SI{450}{\nano\meter} with over 2000 comb lines, far exceeding the telecommunications E-, S-, C-, L-, and U- bands at \SI[number-unit-product={\text{-}}]{29.6}{\giga\hertz} spacing. 
This ultra-broad comb is realized with gross on-chip power consumption below \SI{7}{\watt}, including contributions from both the microwave pump and the laser diode current.
Moreover, a low microwave pump power of \SI{13}{\dbm} gives rise to a \SI[number-unit-product={\text{-}}]{80}{\nano\meter} comb span, representing a 16-fold power reduction compared to non-resonant electrodes. 
This advancement sets the stage for further system-level integration with low-power, analog CMOS-compatible microwave circuitry as well as efficient \ch{GaAs} and \ch{GaN} MMICs for hybrid microwave photonic processors and broadband sensors.
Beyond comb generation, the triply resonant Hamiltonian given by \cref{eq_The_H} also permits effective interaction between the microwave and optical modes, mediated instead by an optical pump. 
Extending the present integrated architecture to superconducting circuits can thus facilitate quantum state transfer of millimeter-wave superconducting qubits with less stringent cooling requirements \cite{sahuQuantumenabledOperationMicrowaveoptical2022,sahuEntanglingMicrowavesLight2023}.
By replacing the pulley-style optical coupler with a coupling resonator for frequency-selective coupling \cite{hu2022high}, we anticipate that self-referenced octave-spanning EO comb generation may be within reach. 
Such broadband EO combs have the potential to significantly enhance low-noise microwave generation through partial frequency division \cite{sun2024integrated,kudelin2024photonic,zhao2024all,li2014electro}. Notably, the number of comb lines directly impacts the extent of phase noise suppression, potentially achieving more than 60 dB in our implementation.
Our results therefore represent a significant step towards the field deployment of EO comb technology, establishing the monolithic microwave co-design strategy for high-performance integrated EO photonics.

%%%%%%%%%%%%%%%%%%%%%%%%%%%%%%%%%%%%%%%%%%%%%%%%%%%%%%%%%%%%%%%%%%%%%%%%%%%%%%%%%%%%%%%%%%%%%%%%%%%%%%%%%%%%%%%%%%%%%
%%%%%%%%%%%%%%%%%%%%%%%%%%%%%%%%%%%%%%%%%%%%%%%%%% Main Content Ends here %%%%%%%%%%%%%%%%%%%%%%%%%%%%%%%%%%%%%%%%%%%
%%%%%%%%%%%%%%%%%%%%%%%%%%%%%%%%%%%%%%%%%%%%%%%%%%%%%%%%%%%%%%%%%%%%%%%%%%%%%%%%%%%%%%%%%%%%%%%%%%%%%%%%%%%%%%%%%%%%%
\medskip
\begin{footnotesize}
\noindent\textbf{Acknowledgments}
We acknowledge early work from (former PhD student) Charles Möhl on using an end-shorted CPW for symmetry-matched triply resonance.
The samples were fabricated in the EPFL Center of MicroNanoTechnology (CMi) and the Institute of Physics (IPHYS) cleanroom. 
This work has received funding from Swiss National Science Foundation grant no. 211728 (Bridge Discovery) and 216493 (HEROIC).
The LTOI wafers were fabricated in Shanghai Novel Si Integration Technology (NSIT) and the SIMIT-CAS.

\noindent\textbf{Author contributions} 
J.Z. and J.R. designed the photonic resonator.
C.D. and G.S.B. designed the microwave resonator.
C.W. and X.O. prepared the LTOI substrate.
C.W. and J.Z. fabricated the devices.
J.Z., G.L., G.S.B., J.H., C.W., W.K., T.B., and N.K. carried out the measurements.
J.Z., J.R., G.S.B., G.L., C.D., W.K., T.B., and N.K. analyzed the data.
J.Z., J.R., G.S.B., W.K., T.B., and T.J.K. wrote the manuscript with contributions from all authors.
Z.L. and  M.C. helped with the project.
J.R., G.S.B., and T.J.K. supervised the project.

\noindent\textbf{Competing interests}
 The authors declare no competing financial interests.

\noindent\textbf{Data Availability Statement} The code and data used to produce the plots within this work will be released on the repository \textit{Zenodo} upon publication of this preprint.
\end{footnotesize}

\bibliography{refs}	
\bibliographystyle{apsrev4-2}
\end{document}

\title{ Supplementary Information: \it{Integrated Triply Resonant Electro-Optic Frequency Comb in Lithium Tantalate}  }

\author{Junyin Zhang}\thanks{These authors contributed equally.}
\affiliation{Institute of Physics, Swiss Federal Institute of Technology Lausanne (EPFL), CH-1015 Lausanne, Switzerland}
\affiliation{Center of Quantum Science and Engineering, EPFL, CH-1015 Lausanne, Switzerland}

\author{Chengli Wang}\thanks{These authors contributed equally.}
\affiliation{Institute of Physics, Swiss Federal Institute of Technology Lausanne (EPFL), CH-1015 Lausanne, Switzerland}
\affiliation{Center of Quantum Science and Engineering, EPFL, CH-1015 Lausanne, Switzerland}

\author{Connor Denney}\thanks{These authors contributed equally.}
\affiliation{Departement of Electrical Engineering, Colorado School of Mines, Golden, Colorado 80401, United States}

\author{Grigory Lihachev}
\affiliation{Institute of Physics, Swiss Federal Institute of Technology Lausanne (EPFL), CH-1015 Lausanne, Switzerland}
\affiliation{Center of Quantum Science and Engineering, EPFL, CH-1015 Lausanne, Switzerland}

\author{Jianqi Hu}
\affiliation{Institute of Physics, Swiss Federal Institute of Technology Lausanne (EPFL), CH-1015 Lausanne, Switzerland}
\affiliation{Center of Quantum Science and Engineering, EPFL, CH-1015 Lausanne, Switzerland}

\author{Wil Kao}
\affiliation{Institute of Physics, Swiss Federal Institute of Technology Lausanne (EPFL), CH-1015 Lausanne, Switzerland}
\affiliation{Center of Quantum Science and Engineering, EPFL, CH-1015 Lausanne, Switzerland}

\author{Terence Blésin}
\affiliation{Institute of Physics, Swiss Federal Institute of Technology Lausanne (EPFL), CH-1015 Lausanne, Switzerland}
\affiliation{Center of Quantum Science and Engineering, EPFL, CH-1015 Lausanne, Switzerland}

\author{Nikolai Kuznetsov}
\affiliation{Institute of Physics, Swiss Federal Institute of Technology Lausanne (EPFL), CH-1015 Lausanne, Switzerland}
\affiliation{Center of Quantum Science and Engineering, EPFL, CH-1015 Lausanne, Switzerland}

\author{Zihan Li}
\affiliation{Institute of Physics, Swiss Federal Institute of Technology Lausanne (EPFL), CH-1015 Lausanne, Switzerland}
\affiliation{Center of Quantum Science and Engineering, EPFL, CH-1015 Lausanne, Switzerland}

\author{Mikhail Churaev}
\affiliation{Institute of Physics, Swiss Federal Institute of Technology Lausanne (EPFL), CH-1015 Lausanne, Switzerland}
\affiliation{Center of Quantum Science and Engineering, EPFL, CH-1015 Lausanne, Switzerland}

\author{Xin Ou}
\affiliation{State Key Laboratory of Materials for Integrated Circuits, Shanghai Institute of Microsystem and Information Technology, Chinese Academy of Sciences, Shanghai, China}

\author{Johann Riemensberger}
\email[]{johann.riemensberger@ntnu.no}
\affiliation{Norwegian University of Science and Technology, NTNU, NO-7491 Trondheim, Norway}

\author{Gabriel Santamaria-Botello}
\email[]{gabriel.santamariabotello@mines.edu}
\affiliation{Departement of Electrical Engineering, Colorado School of Mines, Golden, Colorado 80401, United States}

\author{Tobias J. Kippenberg}
\email[]{tobias.kippenberg@epfl.ch}
\affiliation{Institute of Physics, Swiss Federal Institute of Technology Lausanne (EPFL), CH-1015 Lausanne, Switzerland}
\affiliation{Center of Quantum Science and Engineering, EPFL, CH-1015 Lausanne, Switzerland}

\maketitle

\renewcommand{\thefigure}{S\arabic{figure}}

\tableofcontents
\section{Cavity electro-optics and comb generation}
\subsection{Hamiltonian formalism}
The Pockels effect drives frequency conversion processes in the lithium tantalate (\LT) microresonator forming the comb generator. 
It alters the optical permeability $\eta = 1/\epsilon_r$ in response to a microwave electric field $E^\mathrm{m}$ such that
\begin{equation}
    \Delta \eta_{ij} = r_{ijk} E_k^\mathrm{m},
\end{equation}
where $r_{ijk}$ and $\epsilon_r$ denote the Pockels tensor components and relative permittivity, respectively. 
Here we use the Einstein notation.
According to Bethe--Schwinger perturbation theory for cavities \cite{bethePerturbationTheoryCavities1943}, the change in optical impermeability, in turn, implies a change in the refractive index $n$ 
\begin{equation}\label{eq:cavity_perturbation}
    \Delta n = \frac{n}{2 \epsilon_0} \frac{ \iiint D_{i}^\mathrm{opt} \Delta\eta_{ij} D_j^\mathrm{opt}\,d\mathcal{V} }{ \iiint E_{i}^\mathrm{opt} \epsilon_{ij} E_j^\mathrm{opt}\,d\mathcal{V} },
\end{equation}
where $D_{i}^\mathrm{opt}$ and $E_{i}^\mathrm{opt}$ correspond to the optical electric displacement and field components, and $\epsilon_0$ and $\epsilon_{ij}$ denote the vacuum permittivity and the permittivity tensor components at optical frequencies.
% The phase retardation of optical waves by an applied electric field can be translated to amplitude modulation by wrapping the optical waveguide into a cavity, as the interference of optical waves during several round trips transform the phase shift into a frequency shift.
Consider first just one single optical cavity mode.
The modification of the optical path length of the cavity then leads to a shift in its resonance frequency.
The parametric dependence of the resonance frequency on $E^\mathrm{m}$ is encapsulated in the interaction Hamiltonian
\begin{equation}\label{eq:pockels_hamiltonian}
    \hat{\mathcal{H}}^\mathrm{EO} = - \frac{1}{\tau} \hbar \hat{\phi} \hat{a}^\dagger \hat{a},
\end{equation}
where $\hat{a}$ and $\hat{a}^\dagger$ are the annihilation and creation operators of the optical field, $\hbar$ is the reduced Planck constant, $\tau$ is the round trip time of the optical cavity, and $\hat{\phi}$ is the optical phase delay in response to the microwave field \cite{tsangCavityQuantumElectrooptics2010}.
By introducing a microwave resonator in the realization of a cavity electro-optic system \cite{tsangCavityQuantumElectrooptics2010,javerzac-galyOnchipMicrowavetoopticalQuantum2016}, we have
\begin{equation}\label{eq:ceo_phase_delay}
    \hat{\phi} = \frac{2\pi m}{n} \frac{\partial n}{\partial V} V_\mathrm{ZPF} \left( \hat{b} + \hat{b}^\dagger \right)
\end{equation}
for the $m^\mathrm{th}$-harmonic optical mode. 
Here, $\hat{b}$ and $\hat{b}^\dagger$ are the annihilation and creation operators of the microwave mode.
The corresponding zero-point voltage $V_\mathrm{ZPF} = \sqrt{\hbar \omega_\mathrm{m}/(2C)}$, which increases (decreases) with the microwave resonance frequency $\omega_m$ (the capacitance $C$). 
% \hl{Terence to supply a circuit schematic}.
Combining \cref{eq:pockels_hamiltonian,eq:ceo_phase_delay}, we obtain an interaction Hamiltonian that takes on the same form as the canonical two-mode optomechanical Hamiltonian
\begin{equation}\label{eq:two_mode_interaction_hamiltonian}
    \hat{\mathcal{H}}^\mathrm{EO}
    =
    -\hbar g_0' \hat{a}^\dagger \hat{a} \left( \hat{b} + \hat{b}^\dagger \right).
\end{equation}
For cavity electro-optics, the vacuum coupling rate 
\begin{equation}
    g_0' = \frac{1}{\tau} \frac{2\pi m}{n} \frac{\partial n}{\partial V} V_\mathrm{ZPF}
\end{equation} 
can be interpreted as the frequency shift of the optical mode induced by a single microwave photon.
Consider instead a triply resonant system with two optical modes ($\hat{a}_1$, $\hat{a}_2$) and one microwave mode ($\hat{b}$).
\Cref{eq:cavity_perturbation,eq:two_mode_interaction_hamiltonian} can be extended as
\begin{align}\label{eq:triply_resonant_hamiltonian}
\begin{split}
    \mathcal{\hat{H}}^\mathrm{EO} = &-\hbar g_0'\left(\hat{a}_1^\dagger\hat{a}_1+\hat{a}_2^\dagger\hat{a}_2\right)\left(\hat{b}+\hat{b}^\dagger\right)\\
    &-\hbar g_0\left(\hat{a}_1^\dagger\hat{a}_2+\hat{a}_1\hat{a}_2^\dagger\right)\left(\hat{b}+\hat{b}^\dagger\right),
\end{split}
\end{align}
where both self- and cross-mode couplings are included through $g_0'$ and $g_0$ respectively.
EO comb generation concerns cross-mode coupling involving $\hat{a}_1$, $\hat{a}_2$, and $\hat{b}$.

Besides comb generation, it is interesting to point out that cavity electro-optics can be leveraged for quantum coherent microwave-optical transduction.
With a large coherent drive of amplitude $\alpha$ inside the optical cavity (assumed to be a real number without loss of generality), the two-mode interaction Hamiltonian (\cref{eq:two_mode_interaction_hamiltonian}) can be linearized as
\begin{align}
    \mathcal{\hat{\tilde{H}}}^\mathrm{EO} &= -\hbar g_0' \alpha \left( \hat{a} + \hat{a}^\dagger \right) \left(  \hat{b} + \hat{b}^\dagger \right)\nonumber\\
    &= -\hbar g_0' \alpha \left( \hat{a} \hat{b}^\dagger + \hat{a}^\dagger \hat{b} \right)
    -\hbar g_0' \alpha \left( \hat{a} \hat{b} + \hat{a}^\dagger \hat{b}^\dagger \right).
    \label{eq:om_hamiltonian}
\end{align}
The two terms mediate quantum state transfer and two-mode squeezing between the optical and microwave modes, respectively \cite{hanMicrowaveopticalQuantumFrequency2021}.
The effective coupling rate ($g_0'\alpha$) is enhanced by the intracavity pump field.

\subsection{Comb generation}\label{sec:comb_generation}
The comb generator operates in the regime where the free spectral range $\omega_\mathrm{FSR} = 2\pi/\tau = 2\pi f_\mathrm{FSR}$ is close to the resonance frequency of the microwave cavity $\omega_\mathrm{m}$.
% In contrast to standard cavity optomechanics and electro-optics, a large number of optical modes interact here.
This leads to cascaded interactions between optical modes mediated by the microwave
\begin{equation}\label{eq:three_mode_eo_hamiltonian}
    \hat{\mathcal{H}}^\mathrm{EO}
    =
    \sum_{\mu = -N}^{N-1}
    - \hbar g_{0} \left( \hat{a}_{\mu+1} \hat{a}_\mu^\dagger \hat{b}^\dagger + \hat{a}_{\mu}^\dagger \hat{a}_{\mu-1} \hat{b} \right),
\end{equation}
where the cross-coupling terms from \cref{eq:triply_resonant_hamiltonian} satisfying $\omega_{\mu+1}=\omega_{\mu}+\omega_\mathrm{m}$ is included.
For simplicity, the vacuum coupling rate is assumed to be constant over the comb span.
The full system Hamiltonian with an optical drive and a microwave pump at frequencies $\omega_\mathrm{L}$ (close to mode $\mu=\mu'$) and $\omega_\mathrm{D}$ is given by
\begin{equation}
\begin{gathered}
    \hat{\mathcal{H}} =
    \hbar \omega_\mathrm{m} \hat{b}^\dagger \hat{b} + \sum_{\mu = -N}^{N-1} \left(\hbar \omega_\mu \hat{a}_\mu^\dagger \hat{a}_\mu + \mathcal{\hat{H}}^\mathrm{EO}\right)\\
    + i\hbar \sqrt{\kappa_\mathrm{ex, \mu'}}\left( \hat{a}_\mathrm{in} \hat{a}_{\mu'}^\dagger e^{-i\omega_\mathrm{L} t} - \hat{a}_\mathrm{in}^\dagger \hat{a}_{\mu'} e^{i\omega_\mathrm{L} t} \right) + i \hbar \sqrt{\kappa_\mathrm{ex, m}} \left( \hat{b}_\mathrm{in} \hat{b}^\dagger e^{-i\omega_\mathrm{D} t} - \hat{b}_\mathrm{in}^\dagger \hat{b} e^{i\omega_\mathrm{D} t} \right).
\end{gathered}
\end{equation}
For ease of physical interpretation, we go into the frame co-rotating with microwave mode frequency $\omega_\mathrm{m}$ and optical gross detuning, defined in the main text as $\Delta_{\mu} = D_{\mathrm{int}}(\mu) + \Delta_\mathrm{L}$. 
Here, we account for the integrated dispersion $D_{\mathrm{int}}(\mu) = \omega_\mu - \left[\omega_{\mu'}+(\mu-\mu')D_1\right]$, where $D_1$ represents the spacing of a uniform frequency grid centered at the driven mode $\mu'$.
The associate input laser and microwave pump detunings are in turn denoted as $\Delta_\mathrm{L} = \omega_\mathrm{L}-\omega_\mathrm{\mu'}$ and $\Delta_\mathrm{m} = \omega_\mathrm{D} - \omega_\mathrm{m}$ , respectively.
Setting $\mu' = 0$, we therefore have
\begin{equation}
\begin{gathered}
    \hat{\mathcal{H}} =
    \sum_{\mu} \hbar\Delta_\mu \hat{a}^\dagger_{\mu}\hat{a}_{\mu} - \sum_{\mu = -N}^{N-1}
    \hbar g_{0} \left[ \hat{a}_{\mu+1} \hat{a}_\mu^\dagger \hat{b}^\dagger e^{-i \left( \omega_\mathrm{m} - D_1 \right) t}
    + \hat{a}_{\mu}^\dagger \hat{a}_{\mu-1} \hat{b} e^{i \left( \omega_\mathrm{m} - D_1 \right) t} \right]\\
    + i \hbar \sqrt{\kappa_\mathrm{ex, 0}} \left\{ \hat{a}_\mathrm{in} \hat{a}_{0}^\dagger e^{-i \left[\Delta_\mathrm{L}+\omega_\mathrm{L}-(\omega_0+\mu D_1)\right] t} - \hat{a}_\mathrm{in}^\dagger \hat{a}_{0} e^{i\left[\Delta_\mathrm{L}+\omega_\mathrm{L}-(\omega_0+\mu D_1)\right] t} \right\} + i \hbar \sqrt{\kappa_\mathrm{ex, m}} \left( \hat{b}_\mathrm{in} \hat{b}^\dagger e^{-i\Delta_m t} - \hat{b}_\mathrm{in}^\dagger \hat{b} e^{i\Delta_m t} \right),
\end{gathered}
\end{equation}
Since we have $\omega_\mathrm{m} = D_1$, the time dependence from the phase term in the first line drops off. 
The internal system dynamics can then be obtained using the Heisenberg equations $\frac{d}{dt} \hat{a}_\mu = \frac{1}{i\hbar} \left[ \hat{a}_\mu, \hat{\mathcal{H}} \right] + \frac{\partial \hat{a}_\mu}{\partial t}$
\begin{equation}\label{eq:mode_eom}
\begin{gathered}
    % \begin{aligned}
        \frac{d}{dt} \hat{a}_\mu(t) = \left[-i\Delta_\mu-\frac{\kappa_\mu}{2}\right] \hat{a}_\mu(t) + i g_0 \hat{a}_{\mu+1}(t) \hat{b}^\dagger(t) + i g_0 \hat{a}_{\mu-1}(t) \hat{b}(t) + \sqrt{\kappa_\mathrm{ex, 0}} \hat{a}_\mathrm{in}(t) e^{-i \left[\Delta_\mathrm{L}+\omega_\mathrm{L}-(\omega_0+\mu D_1)\right] t}\delta_{\mu,0},\\
        \frac{d}{dt} \hat{b}(t) = - \frac{\kappa_\mathrm{m}}{2} \hat{b}(t) + \sum_{\mu = -N}^{N-1} i g_0 \hat{a}_{\mu+1}(t) \hat{a}_\mu^\dagger(t) + \sqrt{\kappa_\mathrm{ex, m}} \hat{b}_\mathrm{in}(t) e^{-i \Delta_\mathrm{m} t},
    % \end{aligned}
\end{gathered}
\end{equation}
where $\kappa_\mathrm{ex, \mu}$ and $\kappa_\mathrm{ex, m}$ ($\kappa_\mu$ and $\kappa_\mathrm{m}$) represent the external coupling (total decay) rates of the modes, and $\delta_{\mu,0}$ is the Kronecker delta function.
For a resonant laser input, the rotating frequency of the phase factor in the optical drive term becomes $\Delta_\mathrm{L}+\omega_\mathrm{L}-(\omega_0+\mu D_1) \approx 2\Delta_\mathrm{L} = 0$.
With a strong microwave pump with power $P_\mathrm{D}$ and phase $\phi_\mathrm{D}$, the microwave equation of motion is simplified to
\begin{equation}
    \frac{d}{dt} \hat{b}(t) \approx - \frac{\kappa_\mathrm{m}}{2} \hat{b}(t) + \sqrt{\kappa_\mathrm{ex, m}} \hat{b}_\mathrm{in}(t) e^{-i \Delta_\mathrm{m} t}.
\end{equation}
Without loss of generality, we set $\phi_\mathrm{D} = 0$.
The microwave pump mediates the interaction between the optical modes.
The steady-state mean intracavity microwave photon number can be computed as $n_\mathrm{m} = \sqrt{P_\mathrm{D}/\hbar \omega_\mathrm{D}}$, and the effective EO coupling rate becomes $g = g_0\sqrt{n_\mathrm{m}}$.

Following \cref{eq:mode_eom} and neglecting vacuum fluctuations, we can write the coupled mode equations describing the dynamics of the slowly varying electride field amplitudes as
\begin{equation}\label{eq:comb_coupled_mode}
    \frac{\mathrm{d}}{\mathrm{d}t}a_{\mu}(t) = \left[- i \Delta_{\mu}-\frac{\kappa_\mu}{2}\right]a_{\mu}(t) + ig \left[a_{\mu - 1}(t) + a_{\mu + 1}(t)\right] + \sqrt{\kpe} a_{\mathrm{in}}(t)\delta_{\mu,0}.
\end{equation}
In the presence of strong birefringence-induced mode-mixing between TE and TM modes, $D_\mathrm{int}$ can no longer be described by a polynomial series. The resonances become locally shifted.

\subsection{Comb slope and span}\label{sec:comb_generation}
The comb envelope slope can then be derived from \cref{eq:comb_coupled_mode}.
Assuming $\Delta_{\mu} \approx 0$, we have
\begin{equation}\label{eq:comb_slope_steady}
    \frac{\mathrm{d}}{\mathrm{d}t}a_{\mu}(t) = -\frac{\kappa_\mu}{2} a_{\mu} + ig \left(a_{\mu - 1} + a_{\mu + 1}\right) = 0.
\end{equation}
in the steady state.
For simplicity, let all $\kappa_\mu = \kappa$.
Assuming a constant slope for the comb envelope, i.e., $\hat{a}_{\mu+1}/\hat{a}_{\mu} = \lambda$ for $\mu > 0$ and $\hat{a}_{\mu-1}/\hat{a}_{\mu} = \lambda$ for $\mu < 0$, we obtain
\begin{equation}
    -\frac{\kappa}{2} + i g \left(\frac{1}{\lambda} + \lambda\right) = 0.
\end{equation}
In the regime of strong EO interaction where $g\gg\kappa$, the solution of $\lambda$ reads
\begin{equation}
    \lambda \approx i \left(1 - \frac{\kappa}{4g}\right) \approx i \exp\left(-\frac{\kappa}{4g}\right).
\end{equation}
The power ratio between adjacent comb lines (for $\mu > 0$) is therefore given by
\begin{equation}\label{eq_comb_slope_power}
    \frac{P_{\mu+1}}{P_{\mu}} = |\lambda|^2 \approx \exp\left(-\frac{\kappa}{2g}\right).
\end{equation}

\subsection{Birefringence Limit}
In this section, we derive the comb power drop when the dispersion contribution of the gross detuning $\Delta_{\mu}$ is no longer negligible. 
As discussed in the main text, birefringence-induced mode-mixing is one major reason for such resonance frequency shift in e.g., \LN, imposing limits on the attainable comb span.
When mode-mixing occurs at mode $\mu = \mu_0 > 0$, the induced resonance frequency distortion $\delta_{\mu_0}$ can be at the gigahertz level, which is much larger than the typical cavity loss rate $\kappa$.
Consequently, we can omit the loss term in \cref{eq:comb_coupled_mode} and obtain the steady-state field amplitude as
\begin{equation}\label{eq_comb_birefringence_drop}
    \frac{\mathrm{d}}{\mathrm{d}t}a_{\mu_0}(t) = -i\Delta_{\mu_0} a_{\mu_0}(t) + ig \left[a_{\mu_0 - 1}(t) + a_{\mu_0 + 1}(t)\right]  = 0.
\end{equation}
Since the resonance distortion induces a significant drop in the comb line field amplitude, given as
\begin{equation}
    a_{\mu_0-1} \gg a_{\mu_0 + 1},
\end{equation}
we can approximate \cref{eq_comb_birefringence_drop} as
\begin{equation}
    -i\Delta_{\mu_0} a_{\mu_0} + ig a_{\mu_0 - 1} = 0.
\end{equation}
The power drop at the mode-mixing point $\mu_0$ is therefore
\begin{equation}
    \frac{P_{\mu_0}}{P_{\mu_0-1}} = \left|\frac{a_{\mu_0}}{a_{\mu_0+1}}\right|^2 = \left|\frac{g}{\Delta_{\mu_0}}\right|^2.
\end{equation}
When mode-mixing occurs, several nearby modes typically become distorted. 
Additionally, the effective EO coupling rate will decrease from $g$ to $g_{\mu_0}$, since the hybrid mode does not fully utilize $\chi^{(2)}_{333}$ of \LT, unlike the TE mode.
The combined effect thus leads to a cascaded power drop as
\begin{equation}
    \frac{P_{\mu_0 + N}}{P_{\mu_0}} \approx \prod_{j=0}^{N -1} \frac{P_{\mu_0 + j + 1}}{P_{\mu_0 + j}} \approx  \left|\prod_{j=0}^{N -1 } \frac{g_{\mu_0+j+1}}{\Delta_{\mu_0+j+1}}\right|^2,
\end{equation}
for a mode $N$ FSRs away.

\subsection{Comb existence range}
\begin{figure}[htbp]
    \centering
    \includegraphics[width=0.5\linewidth
    ]{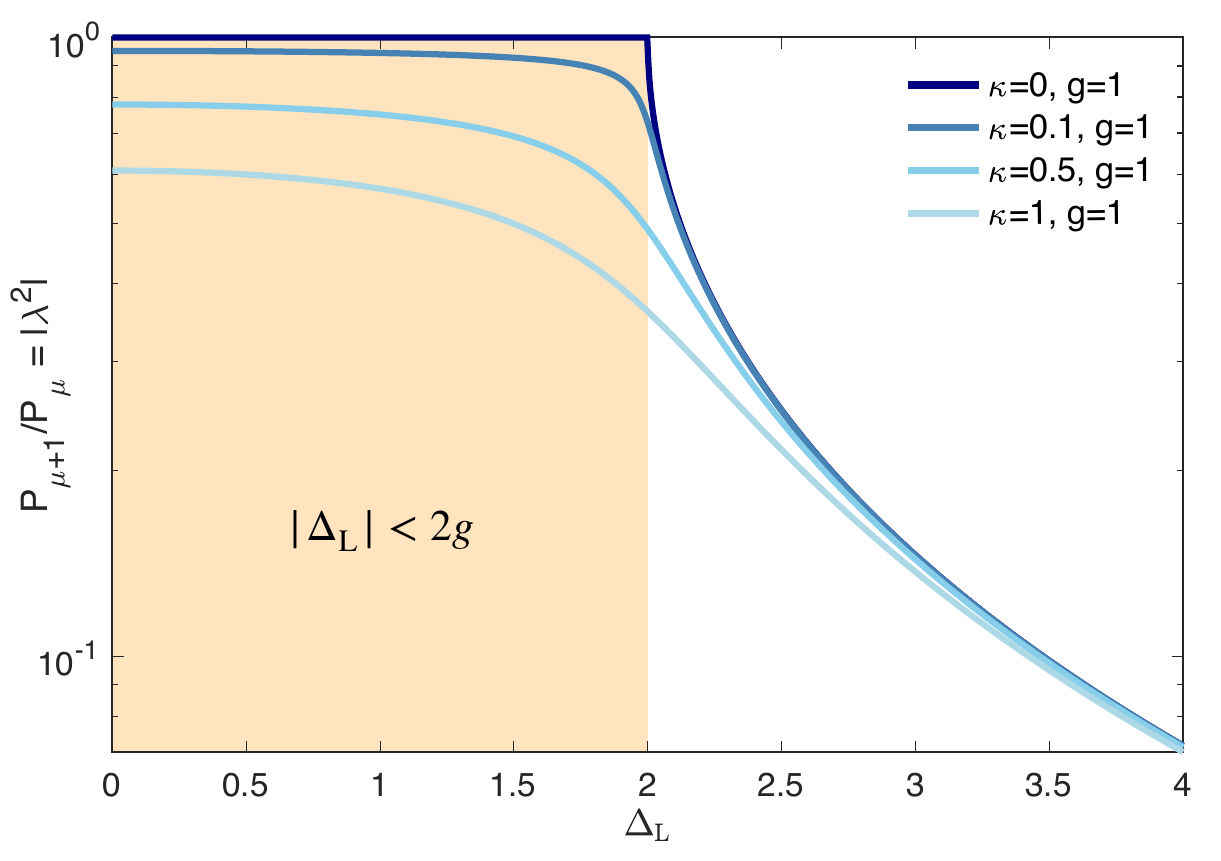}
    \caption{Normalized comb slope $P_{\mu+1}/P_{\mu}$ from simulation, with the same normalized coupling rate \( g \) and different cavity loss rates \( \kappa \). The comb existence range is given by \( |\Delta_{\mathrm{L}}| < 2g \). In the ideal case when \( g \gg \kappa \), the comb slope remains unchanged within the range. The comb slope undergoes a sharp decline when \( \Delta_{\mathrm{L}} = \pm 2g \). This cut-off becomes less sharp as the cavity linewidth \( \kappa \) increases.}
    \label{fig:SI_comb_existance_range}
\end{figure}

In this section, we discuss the comb existence range as a function of laser detuning, which is directly related to the stability of the comb when the input laser frequency fluctuates.
In \cref{eq:comb_coupled_mode}, omitting the \(\kappa\) term for all modes as well as integrated dispersion \(D_{\mathrm{int}(\mu)}\) such that \(\Delta_{\mu} = \Delta_{\mathrm{L}}\), we have
\begin{equation}\label{eq:comb_slope_steady_detuning}
    \frac{\mathrm{d}}{\mathrm{d}t}a_{\mu}(t) = -i\Delta_\mathrm{L} a_{\mu} + ig \left(a_{\mu - 1} + a_{\mu + 1}\right)  = 0.
\end{equation}
Following the constant comb slope assumption as in \cref{sec:comb_generation},
we obtain from \cref{eq:comb_slope_steady_detuning}
\begin{equation}
-i\Delta_\mathrm{L} + i g \left(\frac{1}{\lambda} + \lambda\right) = 0.
\end{equation}
Choosing the solution on physical grounds for $\mu > 0$, we have
\begin{equation}
    \lambda = \frac{1}{2}\left(\frac{\Delta_\mathrm{L}}{g} - \sqrt{\left(\frac{\Delta_\mathrm{L}}{g}\right)^2 - 4}\right).
\end{equation}
As the comb slope $\lambda$ decrease significantly when $\left(\Delta_\mathrm{L}/g\right)^2 - 4 > 0$, the comb cut-off occurs at the detuning
\begin{equation}
    |\Delta_\mathrm{L}| > 2g.
\end{equation}
It can be seen that enhancing $g$ with a microwave resonator extends the comb existence range, approaching the optical resonator FSR in our case. 
The expression for the normalized comb slope $ P_{\mu+1}/P_{\mu} =\left| \lambda \right|^2 $ when $\kappa \neq 0 $ can be derived using the same method. 
The numerical simulation results are shown in \Cref{fig:SI_comb_existance_range}. 
In our triply resonant system, with \SI[number-unit-product={\text{-}}]{37}{\dbm} on-chip microwave power, we have $ \kappa / g \approx 0.013$.

\subsection{Analytic solution and pulse width}
In this section, we derive the rigorous analytical solution for the comb equation \cref{eq:comb_coupled_mode} in steady state. 
Utilizing this analytical solution, we subsequently derive the pulse width in the time domain for the EO comb.
In steady state, we have
\begin{equation}
    \frac{\mathrm{d}a_{\mu}}{\mathrm{d}t} = \left(-\frac{\kappa}{2} - i \Delta_{\mu} \right)a_{\mu} + ig \left(a_{\mu - 1} + a_{\mu + 1}\right) + \sqrt{\kpe}a_{\mathrm{in}}\delta_{\mu,0} = 0\label{eq:comb_coupled_mode_steady}.
\end{equation}
Here $\mu=-N/2,\cdots, N/2$, with $N\to+\infty$ being the total number of modes.
Introducing a new basis in the Fourier domain, we define
\begin{equation}
    A_k = \frac{1}{\sqrt{N}}\sum_{\mu}a_{\mu}e^{\frac{2\pi i}{N}\mu k}
    \label{eq:analytic_reciprocal}
\end{equation}
so that $a_{\mu} = \frac{1}{\sqrt{N}}\sum_{k}A_{k}e^{\frac{-2\pi i}{N}\mu k}$. 
Omitting dispersion such that $\Delta_{\mu} = \Delta_\mathrm{L}$ and rewriting \cref{eq:comb_coupled_mode_steady} with $A_k$, we have
\begin{equation}
    \left(-\frac{\kappa}{2} - i \Delta_{\mathrm{L}}\right)A_{k} + 2ig \cos(\frac{2\pi k}{N})A_{k} + \sqrt{\frac{{\kpe}}{{N}}}a_{\mathrm{in}} = 0,
\end{equation}
where
\begin{equation}
    A_k = \sqrt{\frac{\kpe}{N}}\frac{1}{(\kappa/2 + i\Delta_{\mathrm{L}}) - 2ig \cos(2\pi k/N)}a_{\mathrm{in}}\label{eq:solution_A_k}.
\end{equation}
Using \cref{eq:analytic_reciprocal} and taking the limit $N\to+\infty$, we have
\begin{equation}\label{eq:residual_solution}
    \begin{aligned}
        a_{\mu} &= \sqrt{\kpe}\sum_{k}\frac{1}{(\kappa/2 + i\Delta_{\mathrm{L}}) - 2ig \cos(2\pi k/N)}e^{-\frac{2\pi i}{N}\mu k}a_{\mathrm{in}}\\
        &= \sqrt{\kpe}\int_{-\pi}^{\pi}\frac{\mathrm{d}u}{2\pi} \frac{1}{(\kappa/2 + i\Delta_{\mathrm{L}}) - 2ig \cos u}e^{-i\mu u}a_{\mathrm{in}}\Bigg|_{u=2\pi k/N}\\
        &= \sqrt{\kpe}\oint_{|z|=1}\frac{\mathrm{d}z}{2\pi i}\frac{1}{z}\frac{z^{\mu}}{(\kappa/2 + i\Delta_\mathrm{L}) - ig(z+1/z)}a_{\mathrm{in}}\Bigg|_{z=e^{-iu}}.\\
    \end{aligned}
\end{equation}
We found the contour integral can be calculated by the residue theorem as
\begin{equation}\label{eq:the_solution}
    a_{\mu}=\frac{i\sqrt{\kpe}}{2}\frac{z_0^\mu}{\sqrt{(\Delta_\mathrm{L}+i\kappa/2)^2-4g^2}}a_{\mathrm{in}},
\end{equation}
where $z_0$ is the solution for $(\kappa/2 - i\Delta_\mathrm{L})z - ig(z+1/z)z = 0$ within $|z|<1$ for $\mu>0$ (the solution outside $|z|>1$ for $\mu<0$).
The solution provided in \cref{eq:the_solution} confirms the validity of the ansatz used earlier in \cref{sec:comb_generation}, where $a_{\mu+1}/a_{\mu} = z_0$. Additionally, the splitting of the resonance in the optical transmission is also evident from \cref{eq:the_solution}, as it occurs when $\Delta_\mathrm{L}^2 - 4g^2 = 0$.

Now we consider the pulse in the time domain. 
The Fourier domain introduced previously for $A_k$ in \cref{eq:analytic_reciprocal} is actually the physical time. 
The optical field inside the optical cavity at time $t$ is given by
\begin{equation}
    a(t) = \sum_{\mu}a_{\mu}e^{i\mu\omega_{\mathrm{FSR}}t}.
\end{equation}
Here, \(\omega_{\mathrm{FSR}}\) corresponds to the optical FSR, which is related to the comb repetition rate by \(\omega_{\mathrm{FSR}} = 2\pi f_{\text{rep}}\).
Comparing with \cref{eq:analytic_reciprocal}, we have
\begin{equation}
    a(t) = \sqrt{N}A_k,
\end{equation}
where $k = \omega_{\mathrm{FSR}}tN/(2\pi)$. 
With \cref{eq:solution_A_k}, the pulse of EO comb in the time domain can be written as:
\begin{equation}
    a(t) = \frac{\sqrt{\kpe}}{(\kappa/2+i\Delta_{\mathrm{L}}) - 2ig\cos\omega_{\mathrm{FSR}}t}.
\end{equation}
The pulse power $I(t) = |a(t)|^2$ when $\Delta_\mathrm{L}=0$ is
\begin{equation}
    I(t) \propto \frac{1}{\kappa^2 + 16g^2 \cos^2 \omega_{\mathrm{FSR}}t},
\end{equation}
from which we obtain the pulse full width at half maximum
\begin{equation}
    \Delta t = \frac{\kappa}{2g\omega_\mathrm{FSR}}.
\end{equation}
In our system with $\kappa/(2\pi) = \SI{100}{\mega\hertz}$, $g/(2\pi) = \SI{7.5}{\giga\hertz}$, and $\omega_{\mathrm{FSR}}/(2\pi) = \SI{30}{\giga\hertz}$, we obtain the pulse width $\Delta t = \SI{35}{\femto\second}$.

\subsection{Numerical simulations of comb span and existence range}
\begin{figure}[htbp]
    \centering
    \includegraphics[width=1\linewidth
    ]{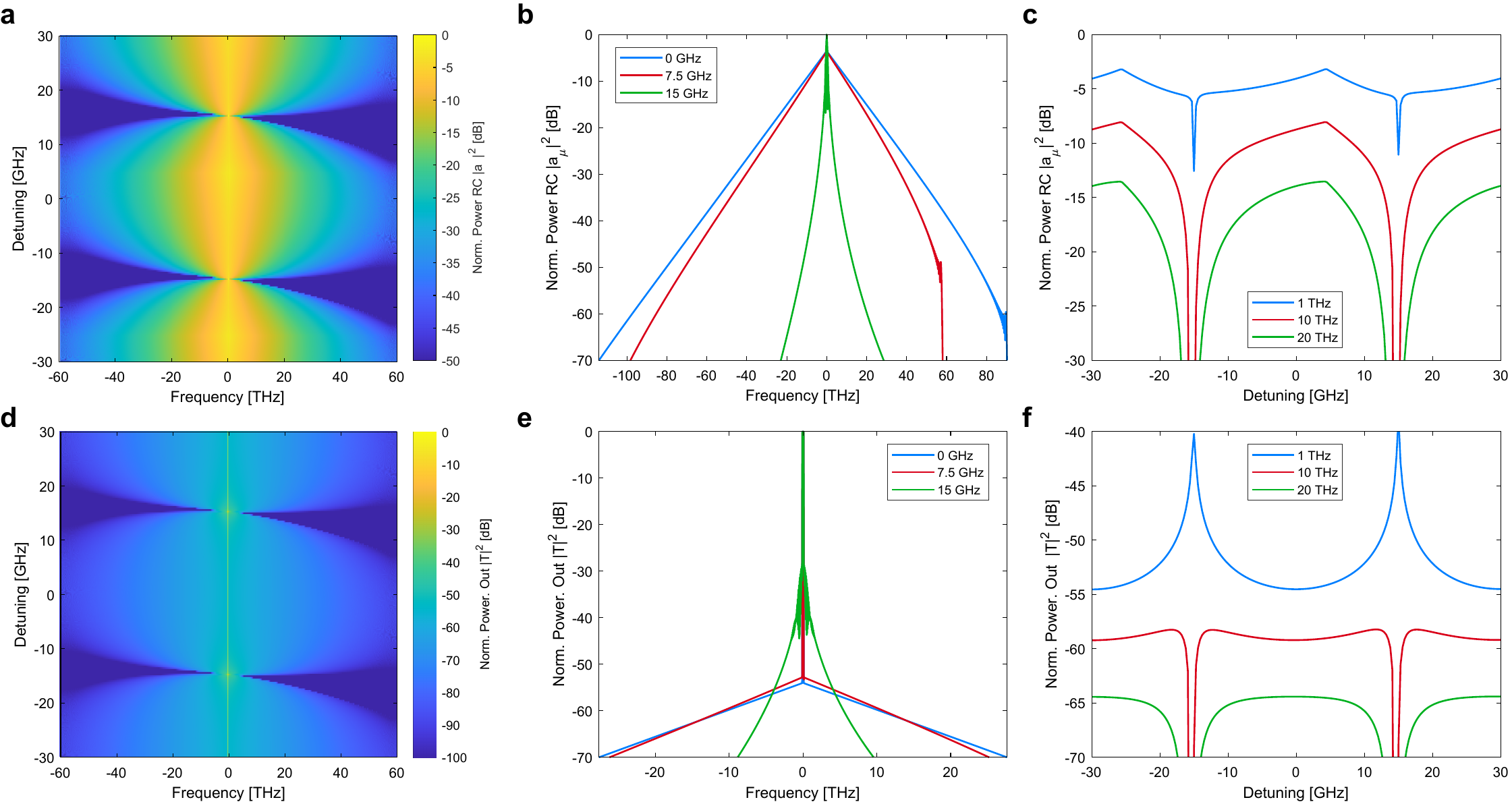}
    \caption{\textbf{Numerical simulation of comb span and existence.}
            (a) Normalized spectrogram of intracavity optical power $\vert a_\mu \vert^2$ as function of laser cavity detuning $\Delta_\mathrm{L}$ at a microwave detuning of $\left(D_1 - \Omega\right)/(2\pi) = \SI{2}{\mega\hertz}$.
            (b) Optical spectra at three different detunings, \SI{0}{\giga\hertz} (blue), \SI{7.5}{\giga\hertz} (red), \SI{15}{\giga\hertz} (green). 
            (c) Comb line power as a function of detuning for comb lines at \SI{1}{\tera\hertz} (blue), \SI{10}{\tera\hertz} (red), \SI{20}{\tera\hertz} (green) positive offset from the pump laser frequency.
            (d,e,f) Same as (a,b,c) but for outcoupled power in the optical bus waveguide.}
    \label{fig:SI_EOcombSim}
\end{figure}

The steady-state solution of the system is found trivially by setting $ \mathrm{d}a_{\mu}/\mathrm{d}t = 0$ and inverting \cref{eq:comb_coupled_mode}. Because we operate the triply resonant system in a regime where the effective EO coupling rate $g=2\pi\times\SI{7.5}{\giga\hertz}$ is comparable to the optical microresonator FSR, we have to include the coupling of the laser to multiple longitudinal modes in the optical microresonator. This regime mandates the use of an explicit time-domain model (Ikeda map) of the field propagation inside the racetrack resonator for the simulation of nonlinear frequency comb generation. For the linear system of the EO comb, we can simply calculate the comb generation for all the resonator modes close to the pump separately and coherently add the generated sideband amplitudes to capture the mutual interferences of the combs generated by coupling the laser field into multiple resonator modes, given by
\begin{equation}
    a_{\mu}  = \sum_{\nu = -3..3} \dfrac{ig \left(a_{\mu - 1} + a_{\mu + 1}\right) +\sqrt{\kpe} a_{\mathrm{in}} \delta_{\mu,\nu}}{-\kappa/2 - i \Delta_{\mu} - i\nu D_1}.
\end{equation}
\Cref{fig:SI_EOcombSim} depicts a calculation of the generated combs inside the racetrack resonator and in the bus waveguide for a free-spectral range $D_1/(2\pi)$ of \SI{30}{\giga\hertz}, internal $\kappa_0/(2\pi)$ and external $\kpe/(2\pi)$ photon decay rates of \SI{30}{\mega\hertz}, a microwave detuning $\Delta\Omega/(2\pi)$ of \SI{2}{\mega\hertz}, and a small normal dispersion $D_2/(2\pi)$ of \SI{2}{\kilo\hertz} for an integrated dispersion of $D_\mathrm{int} = D_2/2\times\mu^2$. Plotting the resulting output power spectrum in \cref{fig:SI_EOcombSim}a and \ref{fig:SI_EOcombSim}d, we find excellent agreement with the observed detuning spectrogram depicted in Fig.~4 of the main manuscript. If the laser is tuned to the center between two resonances, we observe a destructive interference between the generated sidebands at a high offset frequency with a slight tilt that is related to the small microwave detuning. \Cref{fig:SI_EOcombSim}c and \ref{fig:SI_EOcombSim}f depict the change in comb sideband powers as we tune the laser over a full FSR. We find that at an intermediate optical frequency offset of \SI{10}{\tera\hertz}, the comb sideband stays almost constant, and in general, we can tune the laser by more than \SI{15}{\giga\hertz} with only a \SI{1.5}{\decibel} sideband modulation amplitude.

%\hl{Mode splitting, if we are going to present data.}

\section{Microwave engineering for electro-optic frequency comb generation}\label{sec:microwave_engineering}
\subsection{Definitions}

The span of a triply resonant EO comb scales with $\sqrt{\Gcoop}$, where the cooperativity $\Gcoop=4g_0^2n_\mw /\kappa^2$. 
Equivalently defined as in \cref{sec:comb_generation}, 
\begin{equation}
    n_\mw = \frac{4{P_\mw}}{\hbar\omega_\mathrm{\mw}^2}\Qmwo \eta(1-\eta)%\frac{q}{(1+q)^2}
\end{equation}
represents the microwave intracavity photon number, where $P_\mw$ and $\omega_\mw$ denote the input microwave power and frequency, respectively. 
The coupling efficiency $\eta = \Qmwo/(\Qmwo+\Qmwe)$ is related to its intrinsic and external quality factors ($\Qmwo$ and $\Qmwe$, respectively). 
%The coupling ratio $q=\Qmwo/\Qmwe$ of the microwave cavity is related to its intrinsic and external quality factors ($\Qmwo$ and $\Qmwe$). 
The intrinsic and external loss rates of the optical modes are denoted as $\kpo$ and $\kpe$, respectively. 
The total optical cavity loss rate is therefore $\kappa=\kpo+\kpe$.
The cross-mode vacuum coupling rate in \cref{eq:three_mode_eo_hamiltonian} can be computed as
\begin{equation}
    g_0=\frac{\varepsilon_0}{4}\sqrt{\frac{\hbar\omega_\mw\omega_\opt\omega_\sid}{W_\mw W_\opt W_\sid}}\int_\text{\LT} \chi^{(2)}_{ijk} E_{\opt,i} E_{\sid,j}^* E_{\mw,k}\,\mathrm{d}V. \label{eq:SI:g}
\end{equation}
The three-wave mixing process here involves a microwave pump field $\mathbf{E}_\mw$ of frequency $\omega_\mw$, mediating interaction between two optical modes ($\mathbf{E}_{\opt,\sid}$) such that $\omega_\sid=\omega_\opt\pm\omega_\mw$.
The mode fields are normalized to their respective energies using 
\begin{equation}
W_{\mw,\opt,\sid}=\frac{\varepsilon_0}{2}\int \varepsilon_{r}(\omega_{\mw,\opt,\sid}) \abs{\mathbf{E}_{\mw,\opt,\sid}}^2\,\mathrm{d}V,
\end{equation}
where $\varepsilon_{r}(\omega_{\mw,\opt,\sid})$ is the space-dependent relative permittivity of the medium at the corresponding mode frequency. 

\subsection{Microwave resonator design}

Three parameters can be controlled from a microwave engineering standpoint to maximize $\Gcoop$ and thus, the comb span, for a given $P_\mw$: (i) the overlap integral of the vacuum coupling rate, which must be maximized via the engineering of the field distribution, (ii) the external quality factor $\Qmwe =\Qmwo$ for critical coupling and maximization of $n_\mw$, and (iii) the microwave intrinsic quality factor $\Qmwo$ which shall be maximized. 

\subsubsection{Maximize $g_0$ with optimal microwave field distribution}
\label{sec:integral}

\begin{figure}[h]
    \centering
    \includegraphics[width=0.3\linewidth]
    {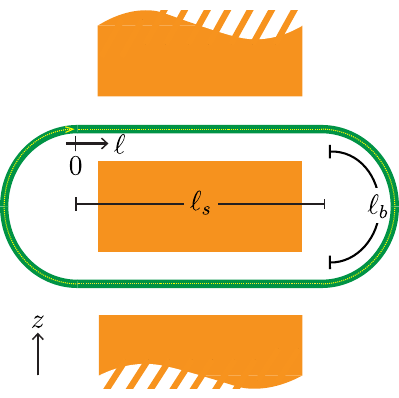}
    \caption{Schematic of racetrack resonator embedded in coplanar waveguide (CPW).}
    \label{fig:SI:CPW_geom_z}
\end{figure}

We focus our study on racetrack resonators where the straight arms of length $\ell_s$ are orthogonal to the optical axis of the crystal ($\hat{c}=\pm \hat{z}$), as depicted in \cref{fig:SI:CPW_geom_z}. A given coplanar waveguide (CPW)-like microwave field distribution permeates the straight arms, and no microwave field exists in the bent arms of arc length $\ell_b$. A longitudinal coordinate $\ell$ is defined along the entire perimeter of the racetrack resonator.

The overlap integral in \cref{eq:SI:g} is computed inside the optical waveguide and within the straight sections only, where $\chi^{(2)}\neq0$ and $\left|\mathbf{E}_\mw\right| = {E}_\mw\neq0$. 
In this region, we assume the fields are separable in the transversal and longitudinal coordinates, 
i.e., $\mathbf{E}_\mw = \mathbf{\Psi}_\mw(\mathbf{r}_\perp)f_\mw(\ell)$, and $\mathbf{E}_{\opt,\sid}=\mathbf{\Psi}_{\opt,\sid}(\mathbf{r}_\perp)e^{-i\beta_{\opt,\sid}\ell}$, where $\beta_{\opt}$, and $\beta_{\sid}$ are the wavenumbers of the pump and sideband, respectively, and $\int_V\,\mathrm{d}V\equiv\int_S\int_\ell\,\mathrm{d}\ell\,\mathrm{d}S$. 
Then, owing to the similarity between transversal profiles of pump and sideband modes, we consider the $\mathbf{\Psi}_{\opt}$ only, and the expression for the vacuum coupling rate reduces to $g_0=K\times\zeta\times\Delta S^{-1/2}\times I$.
Aside from the pre-factor 
\begin{equation}
K=\chi^{(2)}\left(\frac{\hbar\omega_\mw\omega_\opt\omega_\sid}{2\varepsilon_0{\varepsilon_r(\omega_\opt)\varepsilon_r(\omega_\sid)}}\right)^{1/2},    
\end{equation}
we have an optical filling factor
\begin{equation}
    \zeta = \left(1+\frac{1}{\varepsilon_{r}(\omega_\opt)}\frac{\int_{S-S_\mathrm{wg}}\left|\mathbf{\Psi}_\opt\right|^{2}\,\mathrm{d}S}{\int_{S_\mathrm{wg}}\left|\mathbf{\Psi}_\opt\right|^{2}\,\mathrm{d}S}\right)^{-1}
\end{equation}
that quantifies the fraction of optical energy propagating inside the nonlinear material $S_\mathrm{wg}$ ($\zeta\approx 1$ for a highly confined mode), relative to the total area (including the evanescent tail in $S-S_\mathrm{wg}$). 
The relative effective microwave mode area is given by
\begin{equation}
    \Delta S = \int_S \varepsilon_r(\omega_\mw){\abs{\frac{\mathbf{\Psi}_\mw}{\Psi_{\mw 0}}}^2}\,\mathrm{d}S,    
\end{equation}
where 
\begin{equation}
    \Psi_{\mw 0} = \frac{\int_{S_\mathrm{wg}} \Psi_\mw \abs{\Psi_\opt}^2\,\mathrm{d}S}{\int_{S_\mathrm{wg}} \abs{\mathbf{\Psi}_\opt}^2\,\mathrm{d}S}
\end{equation}
is the average microwave electric field inside the waveguide weighted by the optical mode profile. 
This effective field approximates the microwave field amplitude evaluated at the peak of the optical mode. 
While smaller CPW gaps minimize $\Delta S$, they also degrade $\Qmwo$, as well as the optical Q factors if the electrodes disturb the optical evanescent tails. 
Finally, the phase-matching term
\begin{equation}
    %    I = \frac{1}{L}\int_0^L \hat{c}(\ell)\cdot\hat{z}{f(\ell)}e^{i\frac{\omega_\mw}{v_\mathrm{g}}\ell}\,\mathrm{d}\ell\label{eq:SI:I}
    I = \frac{1}{L}\int_0^L \hat{c}(\ell)\cdot\hat{z}u(\ell)e^{i\frac{\omega_\mw}{v_\mathrm{g}}\ell}\,\mathrm{d}\ell\label{eq:SI:I}
\end{equation}
must be optimized by choosing an appropriate longitudinal microwave field distribution $u(\ell)=\frac{f(\ell)}{\sqrt{\int_0^L\abs{f(\ell')}^2\,\mathrm{d}\ell'}}$. The unit vector $\hat{c}$ is in the direction of the optical axis. The exponential term in \cref{eq:SI:I} contains the group velocity $v_\mathrm{g}$ of the optical modes, and is a consequence of the first-order approximation $\beta_\sid-\beta_\opt \approx \omega_\mw\eval{\dv{\beta}{\omega}}_{\omega_\opt}=\omega_\mw/v_\mathrm{g}$. By convention, $f(\ell)$ is the microwave field component in the $\hat{z}$ direction. In this case, $\hat{c}$ can be a function of $\ell$ if the crystal is poled, and is always parallel (or anti-parallel) to $\hat{z}$. Since both optical modes have common polarization along the racetrack, no extra sign change is required. The total perimeter of the racetrack resonator is $L=2\ell_s + 2\ell_b$. 

For CPW even modes, the microwave field distribution is symmetric with respect to a plane (magnetic wall) perpendicular to the plane of the racetrack, parallel to its straight arms, and passing through its the center. This allows us to write \cref{eq:SI:I} as

\begin{equation}
    I = \frac{1}{L}\int_0^{\ell_s}  u(\ell)e^{i\frac{\omega_\mw}{v_\mathrm{g}}\ell}\,\mathrm{d}\ell 
    \mp e^{i\frac{\omega_\mw}{v_\mathrm{g}}(\ell_s+\ell_b)}\frac{1}{L}\int_{0}^{\ell_s} u(\ell_s-\ell)e^{i\frac{\omega_\mw}{v_\mathrm{g}}\ell}\,\mathrm{d}\ell\label{eq:SI:I2},
\end{equation}
where we have assumed that the optical axis is uniform along each straight arm, but possibly different. Concretely, $\hat{c}=\hat{z}$ in the top arm, and $\hat{c}=\pm \hat{z}$ in the bottom arm. Taking the minus sign implies bilateral poling of the racetrack, which can be done by applying a high DC voltage to the CPW line terminated in open circuit. Then, we note that for the optical sidebands to be resonant, 
\begin{equation}\label{eq:SI:resonant_condition}
    \omega_\mw = 2\pi M f_\mathrm{FSR} = 2\pi M \frac{v_\mathrm{g}}{L},
\end{equation} 
where $M$ is the integer number of optical free-spectral ranges between the two optical modes. 
Defining the integer $p = \mp e^{i\frac{\omega_m}{v_\mathrm{g}}\frac{L}{2}} = \mp e^{i\pi M}$, the racetrack factor $r=\ell_s/L$, the normalized wavenumber $\xi = 2\pi M r$, and applying a change of variables $x=\ell/\ell_s$ to scale the domain to $0\leq x\leq 1$, we can rewrite \cref{eq:SI:I2} as

\begin{equation}\label{eq:SI:overlap_integral}
    I=\sqrt{\frac{r}{2L}}\int_0^{1} \frac{f(\ell_s x)}{\sqrt{\int_0^1\abs{f(\ell_s x)}^2\,\mathrm{d}x}}g_p(x,\xi) \,\mathrm{d}x,
\end{equation}
where, 

\begin{equation}
g_{p}(x,\xi)=\begin{cases}
        2e^{i\xi/2}\cos\xi(x-1/2), & p=1\\
        i2e^{i\xi/2}\sin\xi(x-1/2), & p=-1\label{eq:SI:I_kernel}
    \end{cases}, \; \text{and}\; \quad
    \begin{tabular}{r|c|c}
        %\cline{2-3}
        & $M$ odd & $M$ even \\
        \cline{1-3}
        No poling & $p = 1$ & $p = -1$ \\
        %\cline{1-3}
        Bilateral poling & $p= -1$ & $p = 1$. \\
    %   \cline{2-3}
    \end{tabular}
\end{equation}

\noindent The overlap integral in \cref{eq:SI:overlap_integral}
% \cref{eq:SI:I_kernel} 
has the form of an inner product between a kernel $g_p(x,\xi)$, and a unity-norm field distribution. 
Therefore, the magnitude of $I$ is maximum when %$\frac{f(\ell_s x)}{\sqrt{\int_0^1\abs{f(\ell_s x)}^2\,\mathrm{d}x}}=A g_p^*(x,\xi)$, where $A$ is a complex amplitude independent of $x$. 
${f(\ell_s x)}= A g_p^*(x,\xi)$, where $A$ is any constant. 
Various conclusions can be drawn from the above results:

\begin{figure}[h]
    \centering
    \includegraphics[%width=1\linewidth
    ]{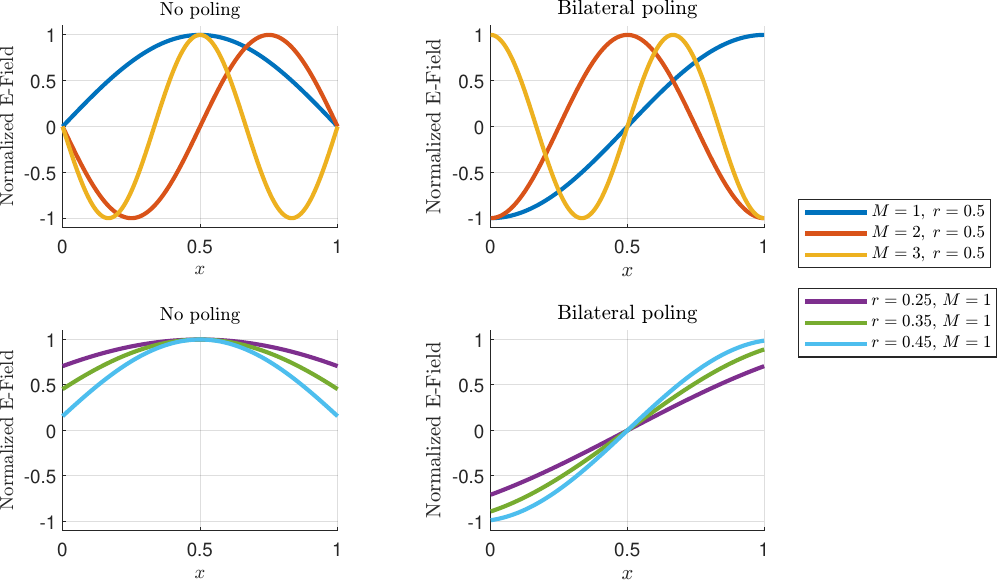}
    \caption{%\textit{
            Optimal microwave field distributions $g_p(x,\xi)$ for different choices of parameters. In the top row, $r=0.5$ is assumed for different values of $M$. In the bottom row, $M =1$ is assumed for different values of $r$. A uniform (left column) and bilaterally poled (right column) optical axis is considered in each case}.
    \label{fig:kernels}
\end{figure}

\begin{figure}[h]
    \centering
    \includegraphics[width=0.8\linewidth
    ]{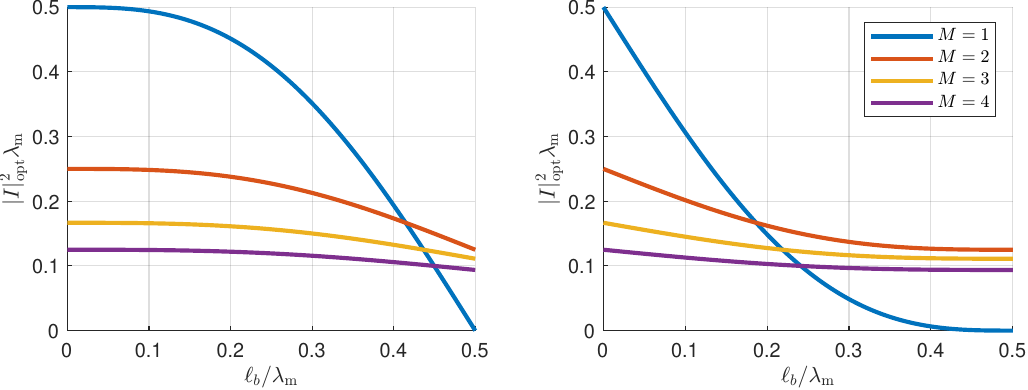}
    \caption{{Maximum $\abs{I}^2\lambda_{\mw}$ versus optical bend length for different choices of $M$. \textbf{Left:} Uniform optical axis. \textbf{Right:} Bilaterally poled optical axis}. 
    }
    \label{fig:Norm_FSR}
\end{figure}

\begin{enumerate}[leftmargin=*]
    \item The kernel $g_p(x,\xi)$ is real-valued except for a global complex factor, for which a standing-wave distribution is optimal. In this case, the value of $\abs{I}^2$ becomes
    \begin{equation}
        {\abs{I}^2}_\mathrm{opt} = \frac{r}{2L}\int_0^1\left|g_p(x,\xi)\right|^2\,\mathrm{d}x = \frac{r}{L}\left( 1+p\frac{\sin(\xi)}{\xi} \right).
        \label{eq:I_opt}
    \end{equation}
    
    \item The optimal standing-wave pattern satisfies $\beta_\mw \ell_s = \xi = 2\pi Mr$.
    From \cref{eq:SI:resonant_condition}, we have $\beta_\mw = \omega_\mw/v_\mathrm{p,\mw} = 2\pi M v_\mathrm{g}/(Lv_\mathrm{p,\mw})$.
    
    Therefore the optimal microwave phase velocity equals the optical group velocity $v_\mathrm{p,\mw}=v_\mathrm{g}$, as is the case for plane waves and whispering-gallery modes. This also implies that the optimal guided microwave wavelength $\lambda_\mw = 2\pi/\beta_\mw = L/M$ is fixed for a given optical waveguide and microwave frequency.
    
    \item For a given $M$ and optical group velocity, $L=M\lambda_\mw$ is fixed, and $\abs{I}^2_\mathrm{opt}$ increases with $r$. In practice, however, there exists a maximum $r<1/2$, or a minimum $\ell_b$, for which optical bending losses are not detrimental.
    One can more easily see this constraint by rewriting the racetrack factor as $r=1/2-\ell_b/L = 1/2-\ell_b/(M\lambda_\mw)$.
    
    \item A consequence of the point above is that choosing a higher $M$ always results in a higher $r$ for a given microwave frequency, since $L$ increases while $\ell_b$ remains constant. However, for a sufficiently small $\ell_{b}/\lambda_{\mw}$, $M=1$ is still optimal because the increase in $L$ has a greater effect on $g_0$ compared to $r$. The full effect can be seen in \cref{fig:Norm_FSR}.
    
    \item For an even CPW mode, the maximum value of $\abs{I}^2 L$ is $1/2$, which occurs for the limiting case $r=1/2$. For a  phase-matched microwave whispering-gallery mode (i.e., $f(\ell)=\mathrm{sgn}(\ell_s-\ell) e^{-i\frac{\omega_\mw}{vg}\ell}$ with no crystal poling) the maximum value is $\abs{I}^2 L=2r=1$ for $r=1/2$ using \cref{eq:SI:I}. Therefore, more complex microwave structures that allow such non-CPW traveling-wave resonances could improve $g_0$ by a factor of $\sqrt{2}$ with respect to the optimal CPW case. Note that if only one of the two straight racetrack arms carried a microwave field (a slot line/ground-signal-type excitation) the optimal solution is again a phase-matched traveling wave, and $\abs{I}^2 L=r/2$. A static solution $f(\ell)=1$ yields $\abs{I}\neq 0$ for $p=1$ only, in which case $\abs{I}^2L=2r \left(\frac{\sin \pi M r}{\pi M r}\right)^2$, which is always lower than \cref{eq:I_opt}. 
\end{enumerate}

\Cref{fig:kernels} shows that the optimal field distribution is cosine-shaped when $p=+1$ (indicating a short-circuit-terminated resonator) 
and sine-shaped when $p=-1$ (indicating an open-circuit-terminated resonator). Note that while the value of $|I|^2L$ for an unpoled, center-fed, half-wavelength, $M=1$ resonator terminated in an open circuit is non-zero, the configuration is sub-optimal. Indeed, the field distribution has a $\abs{\sin\xi(x-1/2)}$ form, and for $r=0.43$ (corresponding to the racetrack resonator measured in this work), \cref{eq:SI:I_kernel} yields
\begin{equation}
    \abs{I}^2L=\frac{r}{2} \frac{
        \abs{\int_0^{1}
            \abs{\sin{\xi (x-1/2)}}
            2e^{i\xi/2}\cos\xi(x-1/2)\,\mathrm{d}x}^2}
    {\int_0^1\abs{\sin{\xi (x-1/2)}}^2\,\mathrm{d}x}
    =0.2537.%0.2705.\; 
    \label{eq:non_ideal_EMW}
\end{equation}
In contrast, for an optimal $\cos\xi(x-1/2)$ distribution, obtainable with a half-wavelength resonator terminated in short circuit, $\abs{I}^2L=0.4978$ %$0.494$ 
as obtained from \cref{eq:I_opt}. 
This is a twofold improvement in cooperativity, or equivalently, a reduction by a factor of 
two in the microwave power requirement for achieving a given comb span.
Note that if the open-circuit-terminated resonator were edge-fed instead of center-fed, the field pattern would become antisymmetric about $x=1/2$, with the form $\sin{\xi(x-1/2)}$, and $\abs{I}^2L=0$, i.e., the vacuum coupling rate $g_0$ would vanish. 
This demonstrates the importance of using the correct topology. 

In practice, there is also a trade-off between phase matching and the physical dimensions of the resonator. Since $r<0.5$ for any practical design, the optimal microwave field distribution has a half-wavelength that is longer than $\ell_s$. Thus, a phase-matched microwave resonator ($v_\mathrm{p,\mw}=v_\mathrm{g}$) will not fit the straight arm of the racetrack as evidenced from the cropped lobes in \cref{fig:kernels} (right). This can be circumvented in two ways: (i) by terminating the resonator in reactive loads using e.g., inductors or bridges to extend the resonator outside the racetrack, or (ii) by using slow-wave structures such that the half-wavelength fits inside the racetrack. The degradation produced by the latter approach on the value of the overlap integral will be small provided that $r\approx0.5$. 
This is the strategy we employ for the present device, although the main motivation is to improve $\Qmwo$, as discussed in \cref{sec:SI:quality_factor}.

\subsubsection{Achieve critical coupling via impedance matching}
The resonant structure must be critically coupled to maximize the electric field enhancement inside the CPW resonator. 
Critical coupling is 
realized when the resonator impedance is matched with external microwave components, typically with a characteristic impedance ($Z_0$) of 50$\Omega$. 
Minimization of microwave reflection is important for integration with high-power devices such as power amplifiers.
A common strategy of impedance-matching a CPW resonator like the one employed in this work is to couple via a reactive element.
This element, either in shunt or in series depending on the topology, serves as a matching network to the source impedance. 
However, such a network is designed for only a given microwave intrinsic Q factor, which may be \textit{a priori} difficult to predict, leading to under-coupling or over-coupling. 
For simplicity, and to maintain design flexibility, we choose to feed these resonators with a ground-signal-ground (GSG) probe positioned off-centered. % the feeding position from the center. 
By adjusting the probe contact position along the CPW, it is possible to achieve critical coupling for a wide range of intrinsic quality factors. 

\begin{figure}[ht]
    \centering
    % \begin{subfigure}[c]{0.45\textwidth}
    %   \begin{subfigure}[c]{\textwidth}
    %       \centering
    %       \includegraphics{fig/SI_offset_scheme.pdf}
    %       \caption{}
 %            \label{fig:SI_offset_scheme}
    %   \end{subfigure}
    %   \begin{subfigure}[c]{\textwidth}
    %       \centering
    %       \includegraphics{fig/SI_offsetFeed.pdf}
    %       \caption{}\label{fig:offset}
    %   \end{subfigure}
    % \end{subfigure}
    % \begin{subfigure}[c]{0.45\textwidth}
    %   \centering
    %   \includegraphics{fig/SI_offsetSmith_edit.pdf}
    %   \caption{}\label{fig:offsetSmith}
    % \end{subfigure}
    \includegraphics[width=0.7\linewidth
        ]{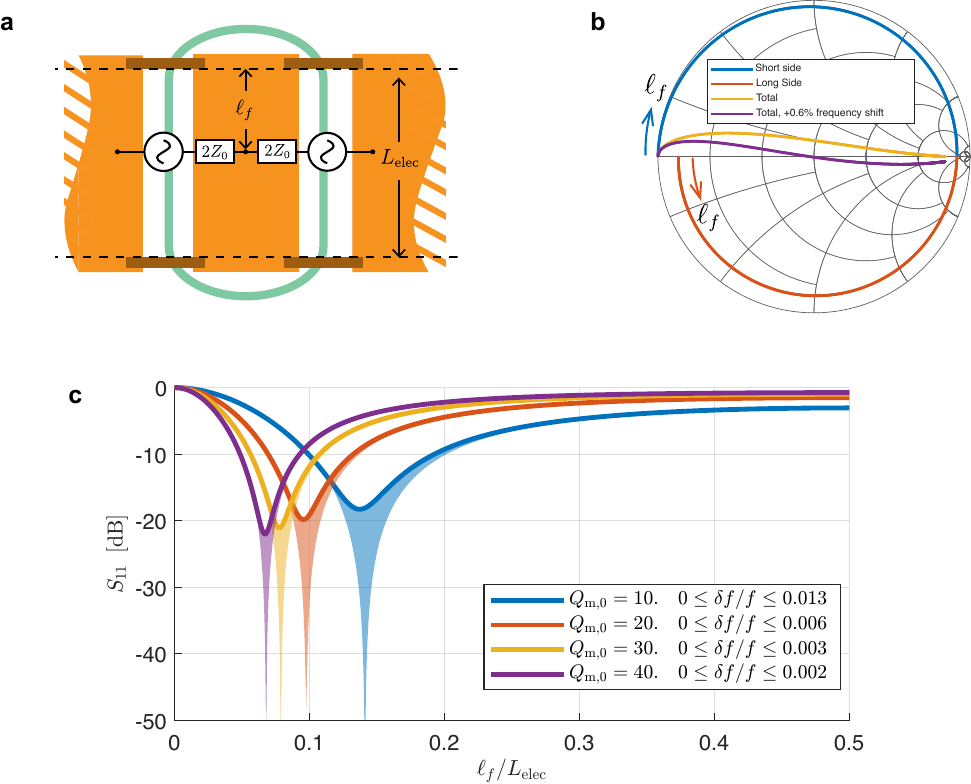}
    \caption{(a) Offset feeding schematic. (b) Smith chart representation of the reflection coefficients as $\ell_f$ is tuned. (c) Reflection coefficient of the resonator as a function of feeding position $\ell_f$ for different choices of intrinsic microwave $\Qmwo$. }
    \label{fig:offsetSmithsf}
\end{figure}

To understand the microwave feeding scheme, consider a $L_\mathrm{elec}=\lambda/2$ transmission line resonator terminated in short circuits, as illustrated in \cref{fig:offsetSmithsf}a. 
Let $\ell_f$ be the distance between one end of the termination and the feeding position. 
In this configuration, the impedance seen by the source is the parallel combination of two short-circuited transmission lines with lengths $\ell_f$ and $L_\mathrm{elec}-\ell_f$.
As $\ell_f$ is swept from $\ell_f=0$ to $\ell_f=L_\mathrm{elec}/2$, the short line becomes longer and the long line becomes shorter. Therefore, as shown in \cref{fig:offsetSmithsf}b, the reflection coefficient seen towards the short side will start at the leftmost side of the Smith chart (a short circuit) and move clockwise in spiral form (non-constant circle due to non-zero losses) until $\ell_f= \lambda/4=L_\mathrm{elec}/2$, at which point the reflection coefficient is the closest to an open circuit. Similarly, the reflection coefficient seen towards the long side will start at a short circuit (since it is $\lambda/2$ away from the short circuit termination) and move counter-clockwise around the Smith chart in spiral form until it is similarly closest to an open circuit. Since at any point on the line both reflection coefficients \textemdash and therefore impedances and admittances\textemdash are nearly complex conjugates of each other, their parallel equivalent is approximately purely resistive, with values that ideally range from $0\,\mathrm{\Omega}$ (at the edge of the resonator) to $\infty \,\mathrm{\Omega}$ (at the center of the resonator), encountering $Z_0$ somewhere in between.

If we consider a half-wavelength CPW resonator short-circuited on either end with characteristic impedance $Z_{0l}$ and intrinsic quality factor $\Qmwo$, the achievable input resistance ranges from %$\abs{Z_L}$ 
$0\,\mathrm{\Omega}$ to $\frac{2}{\pi}\Qmwo Z_{0l}$. Therefore, it is possible in practice to match the resistive part of the input impedance to $50\, \mathrm{\Omega}$ even for low microwave quality factors. The reactive part, however, does not cancel out perfectly for low-Q resonators due to the loss-induced asymmetry of the circle paths in \cref{fig:offsetSmithsf}b, resulting in the yellow path on the Smith chart. This effect is self-corrected with a slightly shifted resonance frequency $f + \delta f$.
The shift moves each reflection coefficient on its circular path in clockwise direction, but at different speeds (the longer side is faster) until complex conjugation is met. This is represented in the purple trace on \cref{fig:offsetSmithsf}b after a $0.6\%$ frequency shift. In practice, the resonance frequency shift at which critical coupling occurs ranges between $0.2\%$ and $1.3\%$ of the nominal resonance frequency for microwave quality factors ranging from $\Qmwo =10$ to $\Qmwo =40$. 
This is shown in \cref{fig:offsetSmithsf}c where the solid lines delineate the reflection coefficient at the nominal frequency for different feeding positions, and the shaded region illustrates the improvement in impedance match with a frequency sweep $\delta f/f$. Here, the reflection coefficient is computed using the equations for the input impedances of two lossy transmission lines in parallel, each terminated with a short circuit:
\begin{equation}
Z_\mathrm{in}^{-1}=Z_{0l}^{-1}\coth\gamma \ell_f+Z_{0l}^{-1}\coth\gamma(L_\mathrm{elec}-\ell_f),    
\end{equation}
where $\gamma$ and $Z_{0l}$ are the propagation constant and characteristic impedance of the line.
Note that our feeding strategy is also valid for open-terminated CPW resonators. However, care must be taken as the phase distribution of the mode varies significantly with the feed position, potentially leading to vanishing modal overlaps.

\subsubsection{Improve microwave intrinsic quality factor via periodic loading of CPW resonator}\label{sec:SI:quality_factor}

\begin{figure}[ht]
    \centering
    % \begin{subfigure}[c]{0.48\textwidth}
    %   \centering
    %   \includegraphics{fig/SI_LayerStack.pdf}
    %   \caption{}
    %   \label{fig:layers}
    % \end{subfigure}
    % \begin{subfigure}[c]{0.48\textwidth}
    %   \centering
    %   \includegraphics{fig/SI_loading.pdf}
    %   \caption{}\label{fig:TCGeo}
    % \end{subfigure}
        \includegraphics{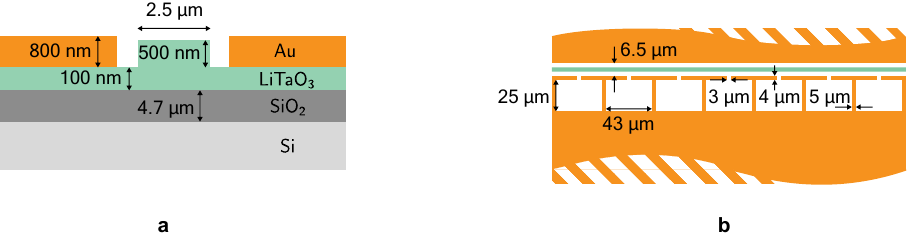}
    \caption{(a) Device layer stack composition (not to scale). (b) Geometry of periodic inductive loading to the narrow CPW line. Only half of the CPW structure is shown.}\label{fig:LayersTCGeo}
\end{figure}

We engineer the CPW resonator to optimize the product $\abs{I}^2\Qmwo$.
We simulate an ungrounded CPW transmission line, with the layer stack composition (geometry) shown in \cref{fig:LayersTCGeo}a (\cref{fig:SI:CPW_geom_z}), using finite-element method (FEM; Ansys HFSS) to extract the characteristic impedance $Z_{0l}$ and propagation coefficient $\gamma=\alpha+i\beta$.
The impedance of the excitation ports are de-embedded via ABCD matrices. 
% A cross-sectional view of the materials and dimensions used in this simulation can be found in \cref{fig:layers}. %\gab{add stack dimensions information (Si, SOW, LT)} 
The CPW electrode gap is chosen to be \SI{6.5}{\micro\meter} to balance the trade-off between modulation efficiency and optical loss.  
The simulation yields $\Qmwo=\beta/(2\alpha)=16.5$ and a microwave effective refractive index $n_{\mathrm{eff}}=c \beta /\omega_\mw=2.228$ at \SI{29}{\giga\hertz}. %Since the modulation depends on E-Field and not voltage, we made the 
%We used ABCD matrices to de-embed the waveguide parameters which resulted in $n_{\mathrm{eff}} = 2.228$ and $Q_{\mu}^{'}=16.5$ at 29GHz. 
The simulated $n_{\mathrm{eff}}$ is close to the optical group index, satisfying the phase matching requirement $\lambda_\mw \approx L$ for $M=1$.  %a good phase match with the optical group velocity, but 
On the other hand, the microwave loss is significant and the electrodes would be too long to fit inside the (optical) racetrack resonator.

Consider a distributed circuit parameter model, where $R'$, $L'$, $G'$ and $C'$ denote the distributed series resistance, series inductance, shunt conductance, and shunt capacitance, respectively. 
{At room temperature}, the conductor Ohmic loss $R'$ limits the $\Qmwo$ of narrow-gap CPWs. 
For $R'\ll \omega L$ and negligible dielectric loss $G'\rightarrow 0$, the propagation coefficient is approximately given by %is mainly limited by conductor loss, characterized by the series distributed resistance $R$ in a distributed $R,L,G,C$ parameter model, where $L$. %When the gap of a CPW gets small, the skin-effect is exaggerated causing the surface current to crowd inside of the edge of the conductor, reducing the effective cross section and increasing loss. In order to improve the Q, we followed \cite{SI:Kharel:21} 
%\con{find and cite T-Cells paper} 
%to implement periodic capacitive loading on our CPW to maintain the E-Field magnitude while decreasing the current crowding in the conductors.
\begin{equation}
    \gamma = \alpha + i\beta = \sqrt{(R'+i\omega L')(G' + i\omega C')}\approx i\omega\sqrt{L'C'} \sqrt{1 -i \frac{R'}{\omega L'}}%\approx \sqrt{LC}\frac{R}{2 L}+i\omega\sqrt{LC}
    \approx \frac{R'}{2 }\sqrt{\frac{C'}{L'}}+i\omega\sqrt{L'C'}. \label{eq:SI:gammaTL}
\end{equation}
The propagation loss $\alpha \propto 1/\sqrt{L'}$ and $\Qmwo \propto L'$ can therefore be improved by increasing the series distributed inductance of the narrow-gap CPW. 
Following the approach of \cite{kharel2021breaking}, we load the line with a periodic series inductance as in \cref{fig:LayersTCGeo}b, synthesized from wide-gap (i.e., large characteristic impedance) short-circuit-terminated slot lines, or T-cells. 
In fact, the microstructure can also be interpreted as capacitive loading of the higher-Q \emph{wide-gap} CPW line. 
In this case, although the microstructure increases $\alpha \propto \sqrt{C'}$, it does not affect the quality factor of the wide-gap transmission line. % is invariant with the capacitive loading. 
We therefore expect $\Qmwo$ of the periodically loaded line to be similar to that of the unloaded wide-gap CPW, with the additional advantage of higher field confinement. 
While periodic loading also introduces a slow-wave effect ($\beta \propto \sqrt{L'C'}$) that leads to a slight phase mismatch, 
the combined effect still results in a twofold cooperativity improvement, as summarized in \cref{tab:WG_Compare}.  
Finally, the microstructure modifies the mode wavelength, allowing us to fit a full $\lambda/2$ resonator along the straight sections of the racetrack resonator.

\begin{table}[h]
    \centering
    \begin{ruledtabular}
    \begin{tabular}{ccccc}
        Transmission line & $\abs{I}^2L$ & $\Qmwo$ & $n_{\mathrm{eff}}$ & Figure of merit: $\abs{I}^2L\Qmwo \propto \mathcal{C}$ \\
        \hline
        %CPW       & 0.217       & 16.5          & 2.228              & 3.580 \\
        Standard CPW       & 0.5       & 16.5          & 2.228              & 8.25 \\
        %T-Cells   & 0.158       & 32.0          & 3.05               & 5.056 \\
        Periodic loading   & 0.48       & 36.7          & 3.05               & 17.61 \\
    \end{tabular}
    \end{ruledtabular}
    \caption{Summary of transmission line simulation results. Although the standard CPW has a better phase-matching overlap with the optical mode, the gain from Q by employing a loaded transmission line leads to a better overall design.}
    \label{tab:WG_Compare}
\end{table}

\begin{figure}
    \centering
    \includegraphics[width=0.6\linewidth
        ]{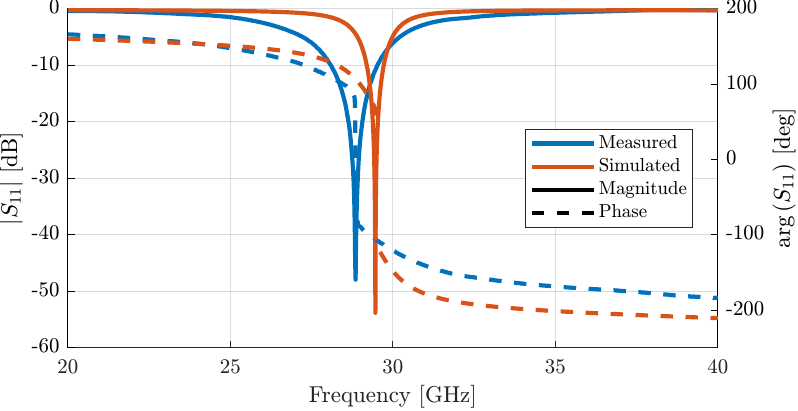}
    \caption{{Measured and simulated $\mathrm{S_{11}}$ of the CPW resonator.}}
    \label{fig:simvsmeas}
\end{figure}

To extract the $\Qmwo$ from the measured and simulated $\mathrm{S}_{11}$ data, we bring the resonator to the critical coupling regime and measure the full width at half maximum. The measured $\Qmwo\approx 12.8$ deviates from the simulated  $\Qmwo\approx 36.7$. % \gab{Add estimated measured vs simulated Q.} 
Other methods such as circle fitting on the Smith chart
% , and complex impedance peak and slope at resonance 
yield similar estimations. 
The significant Q reduction may be attributed to fabrication-induced and piezo-electric loss. 
\Cref{fig:simvsmeas} compares the simulated and measured $\mathrm{S}_{11}$. The difference in resonance frequency therein can be attributed to an underestimation of the \LT dielectric constant in simulation.

\subsection{Comparison between resonant and non-resonant implementations}

While the theory above is useful for comparing different implementations of microwave resonators, a generalization to arbitrary microwave field distributions, whether resonant or not, is possible. Since $\sqrt{\Gcoop} \propto g=\sqrt{n_\mw} g_0$, we can set $n_\mw = W_\mw/(\hbar \omega_\mw)$. As a result,

\begin{equation}\label{eq:SI_final_sqrtG}
    g=\sqrt{n_\mw} g_0 = \frac{\varepsilon_0}{4} \sqrt{\frac{\omega_\opt\omega_\sid}{ W_\opt W_\sid}}\int_\mathrm{\LT} \chi^{(2)}_{ijk} E_{\opt,i} E_{\sid,j}^* E_{\mw,k} \,\mathrm{d}V \approx \frac{\chi^{(2)}_{333}\omega_{\opt}\zeta}{2\varepsilon_r(\omega_\mathrm{p})} \abs{E}_\mathrm{peak}\Lambda,
    % \left|I\right|\sqrt{\int_{0}^{L}\left|f(\ell)\right|^{2}\,\mathrm{d}\ell}, 
\end{equation}
where %$n_p$ is the bulk optical refractive index and 
$\Lambda = \left|I\right|\sqrt{\int_{0}^{L}\left|f(\ell)\right|^{2}\,\mathrm{d}\ell}$. By convention, %in \cref{eq:SI_final_sqrtG} 
we have normalized $f(\ell)$ such that $\max\left[f(\ell)\right]=1$, and $\abs{\Psi_{\mw0}} = \abs{E}_\mathrm{peak}\propto \sqrt{P_\mw}$ is the peak electric field along the resonator, inside the optical waveguide. Higher values of $\abs{E}_\mathrm{peak}$ are achieved with higher-Q microwave resonators that are critically coupled. 
\Cref{eq:SI_final_sqrtG} allows us to compute $g$ from the peak microwave electric field and its geometric distribution along the optical waveguide obtained from driven full-wave simulations excited with available power $P_\mw$.  The factor $\Lambda$ is purely geometrical, with $\Lambda = \abs{I}\sqrt{L_\mathrm{elec}}$ for a (multiple of) half-wavelength standing wave, with $\Lambda \rightarrow 1/2$ in the limiting case where $r\rightarrow 1/2$.  

\begin{figure}
    \centering
    %\includegraphics[%width=1\linewidth
    %]{fig/SI_TLVdistr.pdf}
        \includegraphics[width=0.8\linewidth
        ]{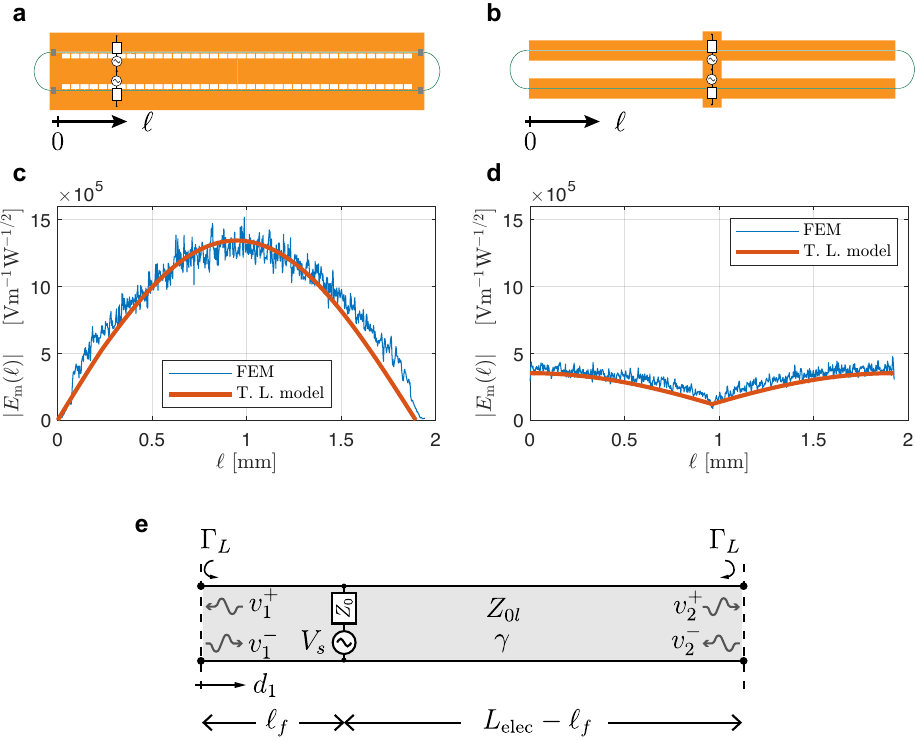}
    \caption{Layout of optical racetrack resonator embedded in microwave resonator (a) and non-resonant electrodes (b). The electric field distribution in resonant and non-resonant topologies is shown in (c) and (d), respectively, and is obtained from the theoretical transmission line model, and compared with FEM simulations. Panel (e) depicts the schematic of the transmission line model.}
    %\label{fig:SI_TLVdistr}
        \label{fig:TLandFields_all}
\end{figure}

\begin{figure}
        \centering
        \includegraphics[width=0.5\linewidth
        ]{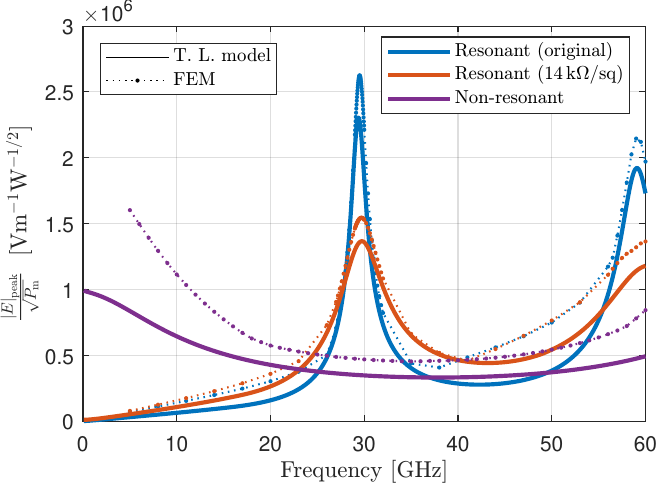}
        \caption{Maximum peak electric field $\abs{E}_\mathrm{peak}$ per square root of available input microwave power inside resonant and non-resonant structures as a function of frequency. }
        \label{fig:Epeak}
    \end{figure}

\begin{figure}
        \centering
        \includegraphics[width=0.5\linewidth
        ]{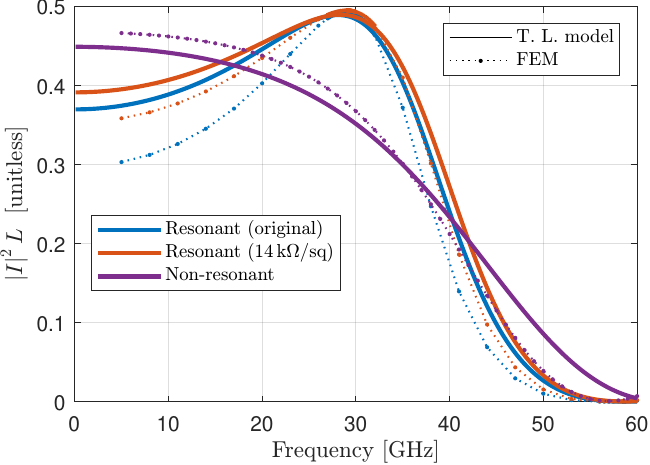}
        \caption{Phase-matching geometric factor  $\abs{I}^2L$ as a function of frequency for resonant and non-resonant structures, calculated with the transmission line model and through FEM simulations.  Note that the non-resonant open-circuited transmission line becomes resonant at $\sim \SI{40}{\giga\hertz}$, at which point $\abs{I}^2L\approx 0.25$. }
        \label{fig:IsqL}
    \end{figure}

\begin{figure}
        \centering
        \includegraphics[width=0.5\linewidth
        ]{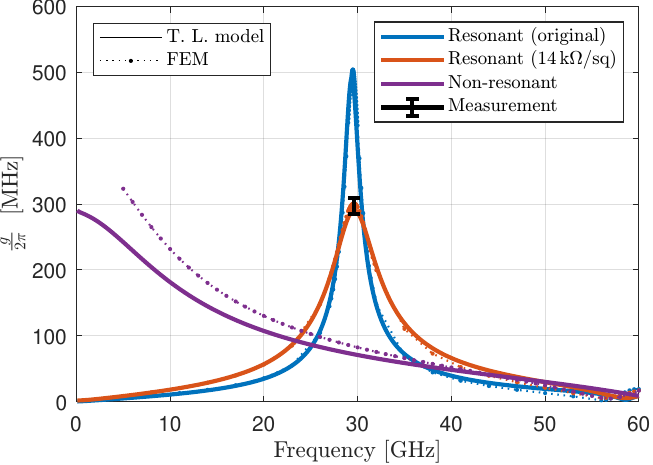}
        \caption{Nonlinear coupling rate $g$ for a microwave available power $P_\mw = \SI{7}{\dbm}$ estimated with transmission line model and with FEM simulations. The measured value of $g$ in the resonator is included. The error bar corresponds to two standard deviations of a sample that comprises seven microwave power levels for the determination of $g$ by means of the comb slope.}%(for the determination of $g_0$).}
        \label{fig:SI_g}
    \end{figure}

Let us compare the performance of the two microwave structures presented in this work, namely, the offset-fed inductively-loaded half-wavelength resonator terminated in short circuits (\cref{fig:TLandFields_all}a), and the center-fed non-resonant pair of \SI{50}{\micro\meter} wide coplanar-strip (CPS) transmission lines terminated in open circuits (\cref{fig:TLandFields_all}b). \Cref{fig:TLandFields_all}c and \cref{fig:TLandFields_all}d show the field distributions in both structures at \SI{29.6}{\giga\hertz} obtained from 3D FEM simulations made in Ansys HFSS, normalized to the square root of the available power.  It is possible to find analytical expressions for the fields in a transmission line of characteristic impedance $Z_{0l}$ terminated with reflection coefficients $\Gamma_\mathrm{L}$.
The line is fed at an arbitrary distance $\ell_f$ from one end by a voltage source with internal real-valued impedance $Z_0$ (or a line with characteristic impedance $Z_0$). Referring to \cref{fig:TLandFields_all}e, the voltage distribution becomes
    
\begin{equation}
        v(d_{1})=\begin{cases}
            v_{1}^{+}\left(e^{\gamma d_{1}}+\Gamma_\mathrm{L}e^{-\gamma d_{1}}\right) & d_{1}\leq\ell_{f}\\
            v_{2}^{+}\left[e^{\gamma\left(L_{\mathrm{elec}}-d_{1}\right)}+\Gamma_\mathrm{L}e^{-\gamma\left(L_{\mathrm{elec}}-d_{1}\right)}\right] & d_{1}\geq\ell_{f}
        \end{cases}\label{eq:SI_Vdstr}.
\end{equation}
The boundary condition at the feed point entirely determines $v_1^+$ and $v_2^+$ through the double equation
    
\begin{equation}
    v_{1}^{+}\left(e^{\gamma\ell_{f}}+\Gamma_\mathrm{L}e^{-\gamma\ell_{f}}\right)=v_{2}^{+}\left[e^{\gamma\left(L_{\mathrm{elec}}-\ell_{f}\right)}+\Gamma_\mathrm{L}e^{-\gamma\left(L_{\mathrm{elec}}-\ell_{f}\right)}\right]=\frac{Z_{\mathrm{in}}}{Z_{\mathrm{in}}+Z_{0}}V_{s},
\end{equation}
where 
\begin{align}
    Z_{\mathrm{in}}={Z_{0l}}\left[\frac{1-\Gamma_\mathrm{L}e^{-2\gamma\ell_{f}}}{1+\Gamma_\mathrm{L}e^{-2\gamma\ell_{f}}}+\frac{1-\Gamma_\mathrm{L}e^{-2\gamma\left(L_{\mathrm{elec}}-\ell_{f}\right)}}{1+\Gamma_\mathrm{L}e^{-2\gamma\left(L_{\mathrm{elec}}-\ell_{f}\right)}}\right]^{-1},
    %\frac{Z_{0l}}{Z_{\mathrm{in}}}=\frac{1-\Gamma_\mathrm{L}e^{-2\gamma\ell_{f}}}{1+\Gamma_\mathrm{L}e^{-2\gamma\ell_{f}}}+\frac{1-\Gamma_\mathrm{L}e^{-2\gamma\left(L_{\mathrm{elec}}-\ell_{f}\right)}}{1+\Gamma_\mathrm{L}e^{-2\gamma\left(L_{\mathrm{elec}}-\ell_{f}\right)}},
\end{align} 
and
\begin{equation}
    \abs{V_\mathrm{s}} = \sqrt{8Z_0P_\mw}.
\end{equation}
The voltage in \cref{eq:SI_Vdstr} is then converted to electric field via 2D FEM simulations of the quasi-TEM mode. Such 2D simulations also provide values for $\gamma$ and $Z_{0l}$.  The transmission line model predicts reasonably well the field distribution along the transmission line for resonant and non-resonant structures when compared to the 3D full-wave FEM simulations as shown in \cref{fig:TLandFields_all}c and \cref{fig:TLandFields_all}d. In \cref{fig:Epeak}, the maximum peak electric fields as a function of frequency, obtained through the transmission line model and FEM simulations are compared. Three cases are studied: the non-resonant open-terminated transmission line and the short-terminated resonator exhibiting intrinsic quality factors $\Qmwo = 36.7$ (expected) and $\Qmwo = 12.8$  (measured). The latter case is modeled in Ansys HFSS by artificially introducing a conductive sheet of resistivity $14\,\mathrm{k\Omega/\square}$ at the silicon-oxide interface. \Cref{fig:IsqL} shows good agreement between the geometric factors of the field profile obtained from the transmission line model and the FEM simulation. From these results, we can estimate $g$ for a given available microwave power $P_\mw$ using \cref{eq:SI_final_sqrtG}, shown in \cref{fig:SI_g} for resonant and non-resonant cases for $P_\mw = \SI{7}{\dbm}$.  In this case, we estimate $g\approx 2\pi\times \SI{290.46}{\mega\hertz}$ for the resonator with $\Qmwo=12.8$. 
 
\begin{figure}[ht]
        \centering
        % \begin{subfigure}[c]{0.48\textwidth}
        %     \centering
        %     \includegraphics{fig/SI_gmeas.pdf}
        %     \caption{}
        %     \label{fig:SI_gmeas}
        % \end{subfigure}
        % \begin{subfigure}[c]{0.48\textwidth}
        %     \centering
        %     \includegraphics{fig/SI_g0meas.pdf}
        %     \caption{}\label{fig:SI_g0meas}
        % \end{subfigure}
        \includegraphics{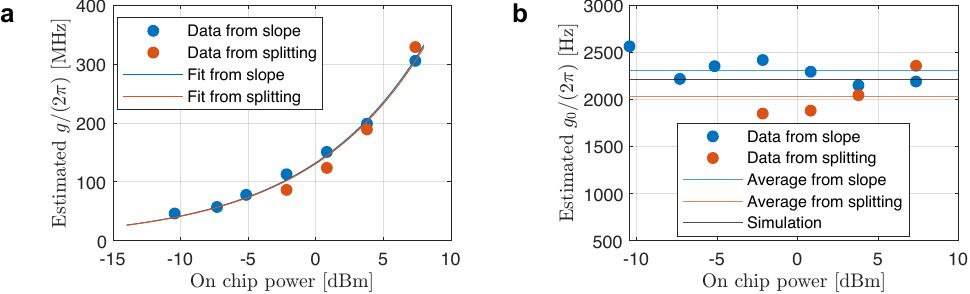}
        \caption{(a) Measurement of the nonlinear coupling rate $g$ via the cooperativity $\sqrt{\Gcoop}$ obtained through the power ratio between consecutive lines of measured comb spectra for different microwave power levels. The lines are single-parameter exponential fits to $g=Ae^{\frac{\ln(10)}{20}P_{\mw,\mathrm{dBm}}}$.  (b) Estimation of the vacuum nonlinear coupling rate $g_0 = g/\sqrt{n_\mw}$.}\label{fig:SI_gg0}
    \end{figure}

In accordance with \cref{sec:comb_generation}, the effective coupling rate $g$ is experimentally measured via two independent methods.
First, we fit the power ratio between consecutive comb lines to the cooperativity as a function of microwave power levels. Second, we measure the optical spectrum for different microwave power levels and extract $g$ from the optical mode splittings. Both methods provide consistent values of $g$ as shown in \cref{fig:SI_gg0}a. 
%This value is consistent with that extracted from optical mode splittings. 
We can then estimate $n_\mw = P_\mw\Qmwo/(\hbar \omega_\mw^2)$ for a critically coupled resonator, resulting in the estimations of $g_0$ shown in \cref{fig:SI_gg0}b. Here, we obtain $g_0/(2\pi) = (2.31\pm0.29) \,\SI{}{\kilo\hertz}$ from slope measurements and $g_0/(2\pi) = (2.03\pm0.47) \,\SI{}{\kilo\hertz}$ from mode splitting measurements, where the quoted uncertainty corresponds to two standard deviations. By extracting the proportionality constant $g/\sqrt{P_\mw}\propto g_0$ via a single-parameter exponential fit of the measurement data in \cref{fig:SI_gg0}a, we obtain $g_0/(2\pi) = (2.22\pm0.04) \,\SI{}{\kilo\hertz}$ from slope measurements and $g_0/(2\pi) = (2.19\pm0.18) \,\SI{}{\kilo\hertz}$ from mode splitting measurements, where the quoted uncertainty corresponds to the 95\% confidence interval of the fit. From simulations and the transmission line model, we obtain $g_0/(2\pi) = 2.20 \,\SI{}{\kilo\hertz}$. 

\section{Dispersion-engineered lithium tantalate waveguides}
\begin{figure}[htbp]
        \centering
        \includegraphics[width=0.8\linewidth
        ]{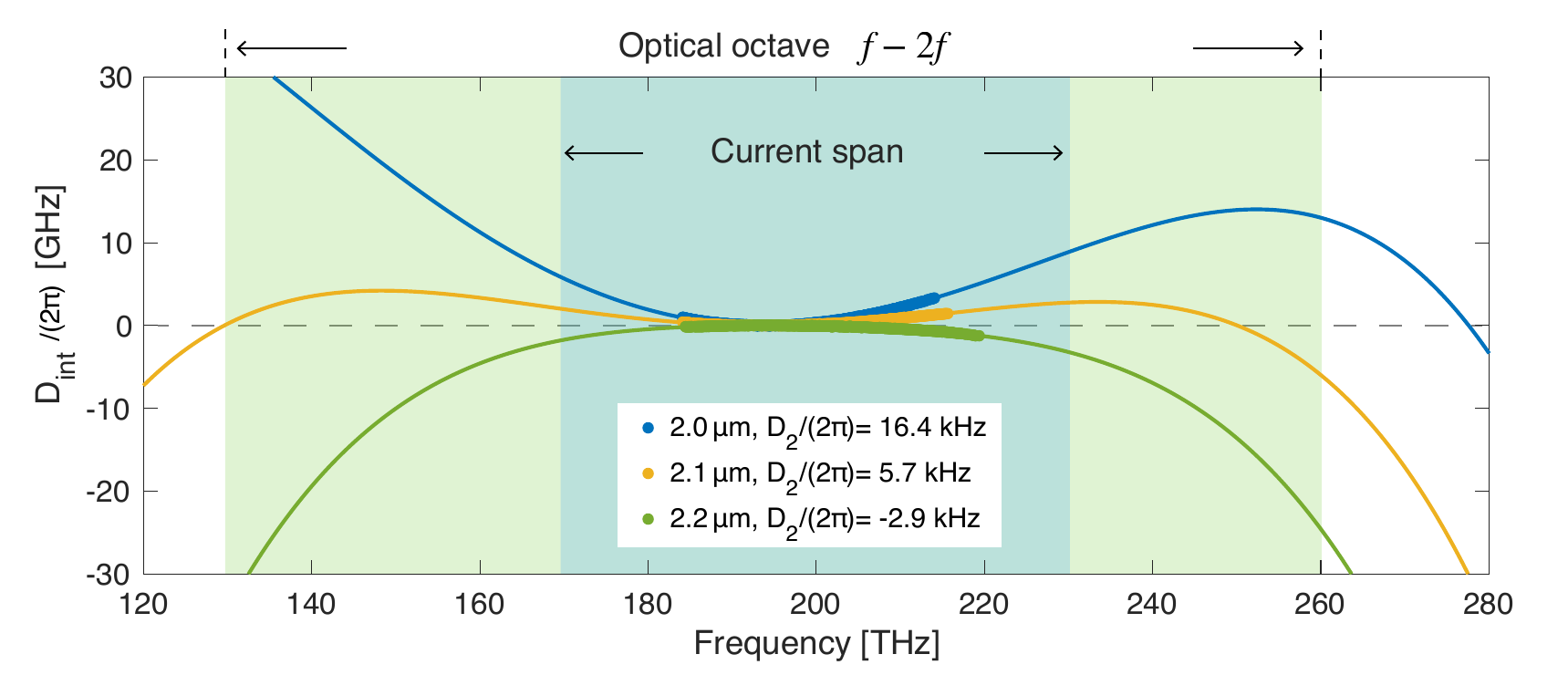}
        \caption{Measured integrated dispersion $D_{\mathrm{int}}/(2\pi)$ for x-cut \LT waveguide with a total thickness of \SI{600}{\nano\meter} (\SI[number-unit-product={\text{-}}]{100}{\nano\meter} slab and \SI[number-unit-product={\text{-}}]{500}{\nano\meter} etch depth), with different waveguide widths.}
        \label{fig:SI_dispersion}
    \end{figure}

In optical resonators, the integrated dispersion $D_{\mathrm{int}}(\mu)$ at the $\mu^\text{th}$ longitudinal mode from the input carrier is defined by
\begin{equation}
    \omega(\mu) = \omega_0 + D_1\mu + D_{\mathrm{int}}(\mu),
\end{equation}
where $D_1 = \omega_\mathrm{FSR}$ is the optical FSR. The integrated dispersion $D_{\mathrm{int}}$ can be expanded as
\begin{equation}
    D_{\mathrm{int}}(\mu) = \frac{1}{2!}D_2 \mu^2 + \frac{1}{3!}D_3\mu^3 + \frac{1}{4!}D_{4}\mu^4 + ...
\end{equation}
To generate an EO comb, the dispersion must fall within the comb existence range such that \( |D_{\mathrm{int}}| < 2g \). 
Consequently, dispersion engineering is essential for ultra-broadband comb generation. 
A unique advantage of integrated photonic waveguides is the engineering degrees of freedom they offer to tailor the dispersion profile. 
In deeply etched photonic waveguides, specifically those used in our work with a \SI[number-unit-product={\text{-}}]{600}{\nano\meter} total thickness (\SI[number-unit-product={\text{-}}]{100}{\nano\meter} slab and \SI[number-unit-product={\text{-}}]{500}{\nano\meter} etch depth), the dispersion profile can be precisely engineered by adjusting the waveguide width. 
This adjustment modifies the \( D_2 \) and \( D_4 \) parameters, allowing them to cancel each other to achieve a flat dispersion.

\section{Optical microresonator characterization}
\begin{figure}[htbp]
        \centering
        \includegraphics[width=1.0\linewidth
        ]{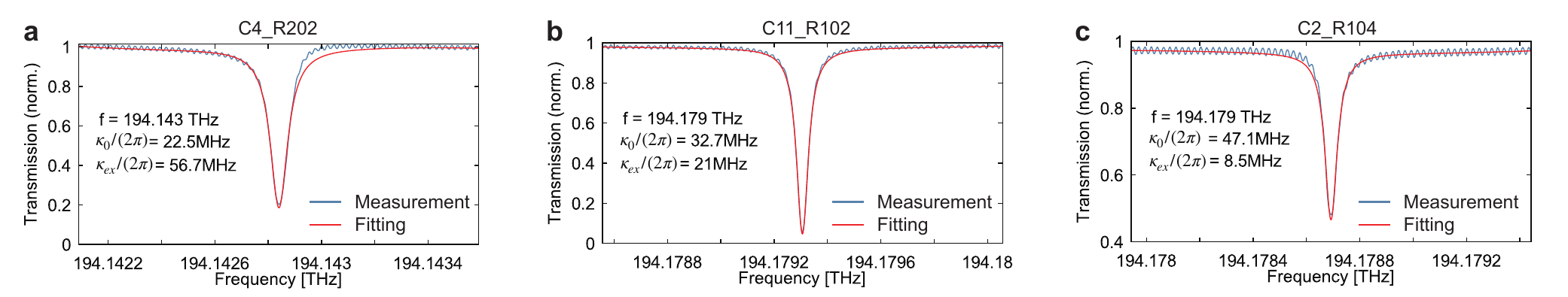}
        \caption{The microresonator resonances used for comb generation. The intrinsic loss rate and external coupling rate are denoted as $\kpo$ and $\kpe$, respectively.}
        \label{fig:SI_characterization}
    \end{figure}
    
The characterization results of the racetrack microresonators used in the experiment are summarized in \cref{fig:SI_characterization}. 
Device \textbf{C4\_R202} (\textbf{C11\_R102}) corresponds to the design with resonant (non-resonant) electrodes in Fig.~3i, generating a \SI[number-unit-product={\text{-}}]{450}{\nano\meter} (\SI[number-unit-product={\text{-}}]{100}{\nano\meter}) comb.
% The device \textbf{C11\_R102} (\cref{fig:SI_characterization}b) is used in Fig.~3i with non-resonant, \SI{50}{\micro\meter} wide microwave electrodes, generating a \SI[number-unit-product={\text{-}}]{100}{\nano\meter} comb.
Device \textbf{C2\_R104}, employed in Fig.~3h, also has resonant electrodes. 
% It is used in Fig.~3h to generate a \SI[number-unit-product={\text{-}}]{100}{\nano\meter} comb. 
In particular, it is under-coupled, resulting in lower comb generation efficiency compared to the over-coupled ones despite a narrower total linewidth. 
This characteristic is desirable for low microwave power-driven comb generation, enabling a \SI[number-unit-product={\text{-}}]{100}{\nano\meter} comb with only \SI{13}{\dbm} of microwave pump.
All the devices have an optical free-spectral range (FSR) of around \SI{29.6}{\giga\hertz} with slight variations. 
In the experiments, the microwave pump frequencies are optimized for each device to match the corresponding optical FSR.

\section{Self-injection locking and hybrid integration}
\subsection{Noise reduction in self-injection locking}
\begin{figure}[htbp]
        \centering
        \includegraphics[width=0.5\linewidth
        ]{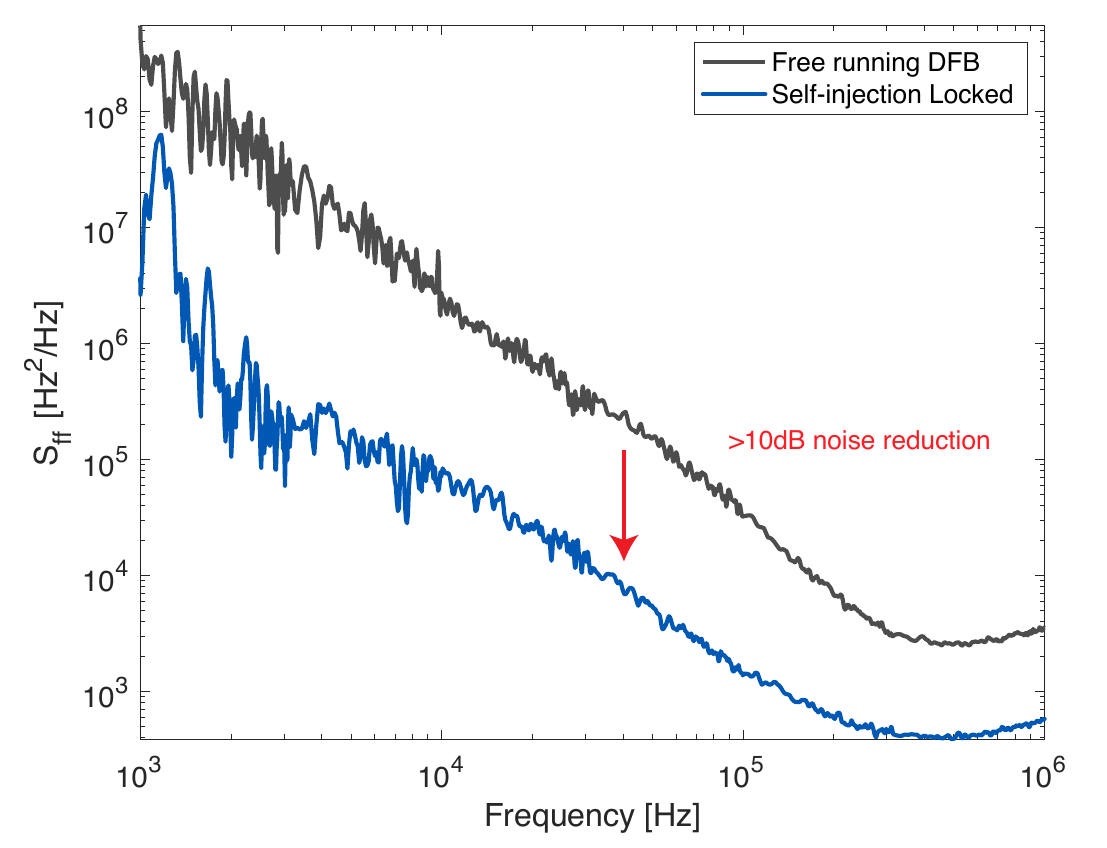}
        \caption{Laser frequency noise of a free-running distributed feedback laser diode (DFB) and when it is self-injection-locked with the \LT resonator, measured using heterodyne detection with a reference laser.}
        \label{fig:SI_SIL_noise}
    \end{figure}

In our system, the comb laser source is a hybrid-integrated distributed feedback (DFB) semiconductor laser diode. 
As shown in \cref{fig:SI_SIL_noise}, the free-running DFB laser exhibits significant frequency noise. 
However, when the DFB laser diode is butt-coupled to the comb generator chip, self-injection locking occurs, resulting in a frequency noise reduction of more than \SI{10}{\dB}.
Self-injection locking involves an external low-loss cavity, where some amount of light is reflected back into the laser via resonant Rayleigh scattering, affording optical feedback to the laser. Consequently, the laser’s linewidth and frequency noise are proportionally reduced, and the reduction is determined by the square of the quality factor of the external resonator $Q_{\mathrm{res}}$\cite{laurent1989frequency,spano1984theory}.
The ultimate limit of this noise reduction would be constrained by the fundamental refractive noise \cite{zhang2023fundamental}.
In our system, a distributed feedback laser (DFB) is self-injection-locked to the racetrack microresonator, also used for comb generation. 
To prevent lock instability, the reflection is provided by only random Rayleigh backscattering from material and fabrication defects in the photonic waveguides and not enhanced by additional reflector structures.
The frequency noise reduction factor can be expressed as \cite{kondratiev2017self}
\begin{equation}
\frac{\delta f}{\delta f_\mathrm{free}} \approx \frac{Q_\mathrm{d}^2}{Q_\mathrm{res}^2}\frac{1}{16\Gamma_\mathrm{res}^2(1+\alpha_g^2)} \propto \frac{1}{Q_\mathrm{res}^2},
\end{equation}
where $\delta f_{free}$ represents the frequency fluctuation of the DFB laser due to drive current noise and temperature fluctuations and $\delta f$ is the frequency deviation of the locked laser. The quality factors of the DFB laser diode and the microresonator are denoted as $Q_\mathrm{d} \sim 10^3\textup{--}10^4$ and $Q_\mathrm{res} \sim 10^6\textup{--}10^7$. 
$\Gamma_\mathrm{res}$ is the resonant amplitude reflection coefficient from the microresonator, and $\alpha_g$ is the phase-amplitude coupling factor, which is approximately unity.

\subsection{Self-injection locking during electro-optic comb generation}
\begin{figure}[htbp]
        \centering
        \includegraphics[width=1.0\linewidth
        ]{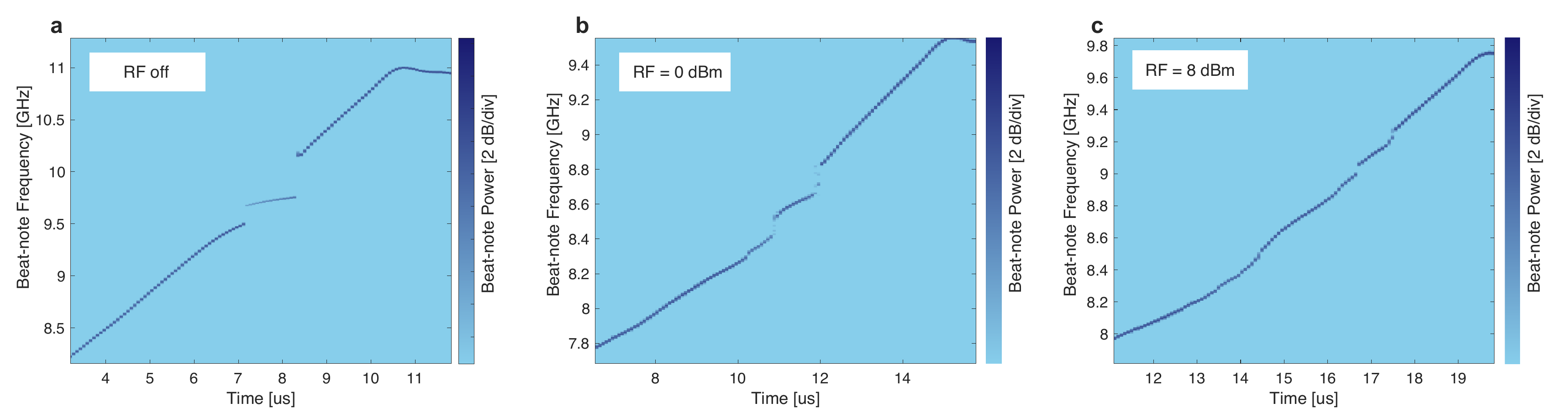}
        \caption{Time-frequency spectrogram of the input carrier mode beat with a reference laser, with different microwave powers applied for comb generation. The microwave frequency matches the optical free spectral range.}
        \label{fig:SI_SIL_range}
    \end{figure}
The significant linewidth reduction due to self-injection locking provides a unique advantage for hybrid-integrated comb generators, as low phase noise is  essential for many applications.
The detuning range $\Delta f_{\mathrm{lock}}$ where self-injection locking can occur depends on the sharpness of the frequency-dependent reflection peak (cavity linewidth of the \LT cavity) and the strength of the optical reflection \cite{kondratiev2017self}, given by
\begin{equation}
    \frac{\Delta f_{\mathrm{lock}}}{f} \approx \sqrt{1 + \alpha_g^2}\frac{\Gamma_{\mathrm{res}}}{Q_\mathrm{d}}.
\end{equation}
\Cref{fig:SI_SIL_range} shows the measured time-frequency spectrogram of the beating between the input carrier mode of the comb and a stable reference laser. Frequency pulling corresponding to self-injection locking can be seen in \cref{fig:SI_SIL_range}a and \ref{fig:SI_SIL_range}b, without and with a weak microwave pump. The optical mode splitting induced by the microwave drive is also observed in \cref{fig:SI_SIL_range}b, as the frequency pulling range is split into two. 
As the microwave pump increases further in \cref{fig:SI_SIL_range}c, no frequency pulling, and thus no self-injection locking, is observed. 
This is due to the reduced reflection $\Gamma_\mathrm{res}$ of the split resonances caused by the applied microwave tone.
In summary, we found that although self-injection locking can coexist with low microwave pump power (narrow comb), the current single-microresonator configuration cannot support the generation of an ultra-broadband EO comb and self-injection locking simultaneously. 
Alternative designs, such as dual-ring comb generation \cite{hu2022high} or adding a separate extended distributed Bragg reflector (E-DBR) section \cite{siddharth2023hertz}, may be required.

\section{Device fabrication}

The devices were fabricated on x-cut single crystalline thin-film \LT wafers from SIMIT-Shanghai. 
The thin-film \LT~wafers were fabricated by ion-cutting and wafer bonding methods. Commercially available optical-grade bulk \LT~were used. Hydrogen ions with an energy of 100 keV and a fluence of 7.0e16 cm$^{-2}$ were implanted into a 4-inch x-cut bulk \LT~wafer, creating an ion-damaged layer beneath the surface. The implanted wafer was then inverted and bonded to a \SI{525}{\micro\meter} thick high-resistivity silicon carrier wafer coated with \SI{4.7}{\micro\meter} thick thermal silicon dioxide. A thermal annealing process \SI{190}{\degreeCelsius} was applied, allowing the separation of the remaining bulk wafer and the exfoliated \LT~thin film. Subsequently, we carried out edge removal of the \LT~thin film and performed chemical mechanical polishing to eliminate the rough and defect-laden layer of \LT~impacted by H-ion implantation, reducing the \LT~film to the target thickness of 600 nm. The wafer stack consists of a 600 nm thin-film \LT, a \SI{4.7}{\micro\meter} thick thermal silicon dioxide, and a \SI{525}{\micro\meter} thick high-resistivity silicon carrier wafer.
The \LT photonics integrated circuits were fabricated using the diamond-like-carbon (DLC) hard mask, which we recently demonstrated for both LNOI and LTOI platforms \cite{wang2023lithium,li2023high}. 
We employed deep-ultraviolet (DUV) stepper photolithography (ASML PAS 5500/350C) to define the photonic waveguides and components. The pattern was transferred into the DLC hard mask layer through oxygen-based dry etching in a reactive ion etcher. Then, the patterns were transferred to the \LT layer using ion-beam etching (Veeco Nexus IBE350). 
Additional chemical cleaning is used to remove the amorphous redeposition of \LT during the ion-beam etching process.
The etch depth is \SI{500}{\nano\meter}, leaving a \SI{100}{\nano\meter} thick slab for dispersion engineering.
The metal layer was patterned with the same DUV tool and fabricated using a lift-off process with silicon dioxide as a sacrificial layer. The silicon dioxide sacrificial layer is removed in buffered oxide etchant after electrode fabrication. 
The metallization layer consists of a \SI[number-unit-product={\text{-}}]{20}{\nano\meter} aluminum adhesion layer and an \SI[number-unit-product={\text{-}}]{800}{\nano\meter} gold layer to reduce Ohmic loss and enhance the microwave quality factor.
After electrode fabrication, \SI{800}{\nano\meter} thick aluminum air bridges were fabricated using a photoresist-based lift-off process (AZ NLOF 2020), with the bridge curvature defined by thermal reflow \cite{reflow_2022}. 
The minimum distance between the air bridges and the photonic waveguide is maintained at more than \SI{1.5}{\micro\meter} to prevent additional optical loss.
Chip singulation is achieved through a combination of dry etching for the \LT and wet oxide layers using fluoride chemistry, as well as deep reactive-ion etching (DRIE) for the silicon carrier \cite{Liu_2021}. 
The process ensures smooth facets for butt coupling with the DFB laser diode.
The residual photoresists are removed using TechniStrip NI555 and cleaned with oxygen plasma.

\bibliography{refs} 
\bibliographystyle{naturemag}